\documentclass[twocolumn,english]{emulateapj}
\usepackage{epsfig}
\usepackage{hyperref}
\usepackage{lmodern}
\usepackage[T1]{fontenc}
\usepackage[latin9]{inputenc}
\usepackage{rotating}
\usepackage{units}
\usepackage{amsmath}
\usepackage{amssymb}

\usepackage{babel}
\usepackage{verbatim}
\usepackage{array}

\newcommand{\rgal}{\mbox{$r_{\rm g}$}}
\newcommand{\rhalo}{\mbox{$r_{\rm h}$}}

\begin{document}
\shortauthors{Brennan et al.}
\shorttitle{Momentum-driven Winds from Black Hole Accretion}

\title{Momentum-driven Winds from Radiatively Efficient Black Hole Accretion \\and Their Impact on Galaxies}

\author{Ryan~Brennan\altaffilmark{1},
  Ena~Choi\altaffilmark{2,1}\footnotemark[*],
  Rachel~S.~Somerville\altaffilmark{1,3},
  Michaela~Hirschmann\altaffilmark{4},
  Thorsten~Naab\altaffilmark{5},
  Jeremiah~P.~Ostriker\altaffilmark{2,6}
}
\footnotetext[*]{E-mail: ena.choi@columbia.edu}

\affil{$^{1}$Department of Physics and Astronomy, Rutgers, The State University of New Jersey, \\136 Frelinghuysen Rd, Piscataway, NJ, 08854, USA\\ 
$^{2}$Department of Astronomy, Columbia University, 550 W 120th Street, New York, NY 10027, USA\\ 
$^{3}$Center for Computational Astrophysics, Flatiron Institute, 162 5th Avenue, New York, NY, 10010, USA\\ 
$^{4}$Sorbonne Universit{\'e}s, UPMC-CNRS, UMR7095, Institut d'Astrophysique de Paris, F-75014 Paris, France\\ 
$^{5}$Max-Planck-Institute for Astrophysics, Karl-Schwarzschild-Str. 1, 85748 Garching, Germany\\ 
$^{6}$Department of Astrophysical Sciences, Princeton University, Princeton, NJ 08544, USA\\
  }

\label{firstpage}

\begin{abstract}

We explore the effect of momentum-driven winds representing radiation
pressure driven outflows from accretion onto supermassive black holes
in a set of numerical hydrodynamical simulations. We explore two
matched sets of cosmological zoom-in runs of 24 halos with masses
$\sim10^{12.0}\--10^{13.4}M_{\odot}$ run with two different feedback
models. Our ``NoAGN'' model includes stellar feedback via UV heating,
stellar winds and supernovae, photoelectric heating and cosmic X-ray
background heating from a meta-galactic background. Our fiducial
``MrAGN'' model is identical except that it also includes a model for
black hole seeding and accretion, as well as heating and momentum
injection associated with the radiation from black hole accretion. Our
MrAGN model launches galactic outflows which result in both
``ejective'' feedback --- the outflows themselves which drive gas out
of galaxies --- and ``preventative'' feedback, which suppresses the
inflow of new and recycling gas. As much as 80\% of outflowing
galactic gas can be expelled, and accretion can be suppressed by as
much as a factor of 30 in the MrAGN runs when compared with the NoAGN
runs. The histories of NoAGN galaxies are recycling-dominated, with $\sim$
70\% of material that leaves the galaxy eventually returning, and the
majority of outflowing gas re-accretes on 1 Gyr timescales without AGN 
feedback. Outflowing gas in the MrAGN runs has higher characteristic 
velocity (500 - 1,000 km/s versus 100-300 km/s for outflowing NoAGN 
gas) and travels as far as a few Mpcs. Only $\sim$ 10\% of ejected 
material is re-accreted in the MrAGN galaxies.

\end{abstract}


\section{Introduction}

It has been estimated by studying absorption lines associated with
active galactic nuclei (AGN) across a large range of luminosities that
upwards of 60\% of AGN exhibit outflows, meaning that outflows are an
important part of the process of supermassive black hole (SMBH) growth
and accretion \citep{Ganguly2008}. This AGN-outflow connection could
even be universal if the typical covering fraction of outflows were
$\sim$ 60\%. Evidence of these outflows being potentially very powerful
comes in the form of observations of broad absorption line quasars,
which make up $\sim20\%$ of the observed QSO population
\citep{Knigge2008}. These systems show evidence of outflows being
launched at velocities as fast as 10,000 km/s or more in the vicinity
of the SMBH \citep{Crenshaw2003, Moe2009, Dunn2010, 
Tombesi2010, Tombesi2013, Tombesi2015}. There is also
evidence for winds driven by AGN in low redshift systems
\citep{Cicone2014, Cheung2016}. However, it is quite difficult to
constrain the size and extent of outflows, and therefore the mass
outflow rate, based on absorption line observations \citep{Arav2012,
  Maiolino2012, Arav2013, Chamberlain2015}. Recently, integral field
unit (IFU) observations have been used to more fully map the gas
around AGN and gain insight into the kinematics of AGN-driven
winds. These studies have revealed that outflows can act on large
scales and entrain large quantities of gas on their way out, although
it is difficult to say for sure how much of a role AGN play in driving
these winds as opposed to stellar feedback processes \citep{Tremonti2007, 
Prochaska2009, Alatalo2011, Rupke2011, Sturm2011, Rupke2013, 
Zakamska2014, Perna2015}.

The overall effect that these outflows have on their host galaxies,
specifically on the gas cycle and future star formation rate, is
unclear, although in general outflows are important for the regulation
of galaxy growth \citep{Lilly2013}. Some observational studies have
investigated the relationship between AGN activity and star formation
in galaxies, but the results have been controversial. Some studies
conclude that winds can remove enough gas to cause negative feedback
on star formation \citep{Rupke2013}, although it is not clear if the
gas will remain outside of the galaxy, and that the presence of a
strong AGN and a suppression of star formation are correlated
\citep{Page2012}. However, other studies find that AGN preferentially
live in galaxies on or even above the star-forming main sequence
\citep{Santini2012, Rosario2013b, Rosario2013} where they can drive
powerful outflows \citep{Genzel2014}, or even that AGN are stimulating
star formation (positive feedback) \citep{Feain2007, Zinn2013}. The
disagreement between different observational studies originates from
the different selection criteria used by various works and the
different timescales on which galaxies and their central supermassive
black holes operate (see \citet{Harrison2017} for details).

On the theoretical side, AGN feedback is often invoked to explain many
observed properties of galaxy populations. It is believed that almost
all galaxies with a bulge component host a supermassive black hole
(SMBH) at their center \citep{Magorrian1998}. The mass of these
central SMBHs is also known to be correlated with several properties
of their host galaxies, specifically the mass, velocity dispersion and
luminosity of their bulge components \citep{Kormendy1995,
  Ferrarese2000, Gebhardt2000, Tremaine2002, Marconi2003,
  Haring2004}. Meanwhile, there is a bimodality in the colors of the
galaxy population \citep{Baldry2004, Bell2004} and the growth of these
two populations suggests that blue, star-forming galaxies are being
transformed into red, quiescent galaxies, with the dearth of galaxies
between the two peaks in color or star-formation rate space indicating
a fast transition timescale for a significant fraction of
transitioning galaxies \citep{Bell2004, Faber2007, Pandya2016}.

With these phenomena in mind (and the large amounts of energy which can be
released by an accreting SMBH \citep{LyndenBell1969}), AGN feedback
becomes an appealing explanation for this galaxy ``quenching'', as
well as for explaining the origin of the various black hole scaling
relations mentioned above. It has also frequently been invoked to
solve the overcooling problem in massive galaxies, suppressing the
late formation of stellar mass and bringing theoretical predictions
into alignment with the observed high-mass ends of the stellar mass
function and stellar mass-halo mass relation \citep[][and references
  therein]{Somerville2015, Naab2016}.

Theorists generally categorize AGN feedback into two broad types:
``radiative'' mode (sometimes called ``quasar'' mode or ``bright''
mode) and ``jet'' mode \citep{Heckman2014}. Radiative mode is
associated with radiatively efficient accretion, relatively high
accretion rates (above a few hundredths of the Eddington rate), and is
thought to be fueled by a classical optically thick, geometrically
thin accretion disc \citep{Shakura1973}. The hard radiation field
emanating from the accretion disc can Compton- and photo-heat as well
as photo-ionize and photo-dissociate gas. In addition, radiation
pressure on dust and free electrons can drive outflows, and this is
likely the origin of the broad absorption line winds discussed above
\citep{Proga2009, Higginbottom2014, Gaskell2016}. When accretion 
rates drop lower than about $\simeq 0.01$ of the Eddington rate, the 
accretion becomes radiatively inefficient, with most of the energy instead 
emerging as highly collimated relativistic jets. These jets are often 
observed at radio frequencies, giving rise to the term ``radio mode''
feedback. The jets appear to be able to heat the diffuse hot halo gas
via giant bubbles (seen in X-ray observations), sound waves, and weak
shocks \citep[see][for reviews]{Fabian2012,Heckman2014}.

Phenomenologically, theorists often speak of ``ejective'' feedback, in
which star formation is quenched due to the removal of the ISM from
the galaxy, and ``preventative'' feedback, in which star formation is
eventually choked off by the lack of fuel, as the inflow of fresh or
recycled gas is suppressed. As noted by \citet{Peng2015} and others,
the implications and effects of these two types of mechanisms for
various aspects of galaxy evolution will differ in important ways. It
is often assumed that radiative mode feedback works in a solely
ejective manner, while jet mode is solely preventative. (As a result,
jet mode feedback is sometimes also referred to as ``maintenance
mode''). Indeed, these assumptions are built into the implementations
of AGN feedback in most semi-analytic models. However, the work
presented here will call into question the first assumption, while
observations of powerful outflows in systems with giant radio jets
seems to challenge the second \citep{Torresi2012}.

Many previous simulations have included prescriptions for AGN
feedback, although usually the radiative mode of feedback is
implemented via deposition of thermal energy, while mechanical
feedback is reserved for lower Eddington ratios and is associated with
the jet mode of feedback \citep{Springel2005b, DiMatteo2012, Dubois2013,
Vogelsberger2014, Hirschmann2014, Khandai2015, Schaye2015, 
Steinborn2015,  Barai2016, Weinberger2017}.
Although such simulations can be successful at reproducing galaxy
properties, it is unclear whether just depositing thermal energy
associated with the radiative mode of black hole accretion can drive
winds similar to those seen in observations. Moreover, thermal energy
input leads to halo X-ray luminosities in disagreement with
observations. If no winds are launched, the thermal energy that is
radiated away is concentrated in the center of the galaxy, resulting
in halo X-ray luminosities that are too low \citep{Bogdan2015}, while
if there are only weak winds, the radiating hot gas that gets pushed
into the halo results in X-ray luminosities that are too high when
compared with observations \citep{Choi2012, Choi2014,Schaye2015}.

The motivation behind the work presented here is that outflowing hot,
shocked gas on scales below those that can be explicitly resolved in
cosmological simulations can impart momentum to gas, which cannot be
radiated away. The physics is similar to the momentum boost occurring
at the end of the Sedov-Taylor phase in a supernova explosion.  Of the
simulations that have modeled AGN feedback, relatively few have
included a momentum-driven prescription associated with radiatively
efficient black hole accretion. Momentum-driven wind scalings
associated with AGN were adopted in the ``Santa Cruz'' semi-analytic
model \citep{Somerville2008}.  \citet{Ostriker2010} examined the effect 
of momentum-driving on the feeding of AGN in a series of one- and 
two-dimension simulations. \citet{Debuhr2010} and
\citet{Debuhr2011} investigated momentum-driven winds from radiation
pressure in hydrodynamic simulations of idealized galaxy mergers.
\citet{Eisenreich2017} investigated the effect of the AGN
prescription used in this work on the metal content of a series of
idealized elliptical galaxies. 
\citet{Hopkins2016} implemented BH feedback via broad absorption
line winds in a similar manner to the mechanical AGN feedback model presented in
\cite{Choi2012} and in the current paper. They investigated the 
interplay between stellar and AGN feedback in isolated galaxies and found
that powerful outflows driven by momentum flux injected via AGN feedback 
strongly suppressed star formation by removing gas. None of these 
hydrodynamical works, however, were in a cosmological context. In terms of 
cosmological simulations, \citet{2017MNRAS.464.2840A} implemented
momentum-driven outflows from AGN in cosmological zoom simulations and
found that they had a strong effect on galaxy and black hole growth.
In addition, the kinetic feedback model of \citet{Weinberger2017} was
shown to bring the baryon content of galaxies in the
\textit{Illustris} simulation into better agreement with observations
\citep{Pillepich2017}, although their mechanical feedback prescription
is associated only with low black hole accretion rates (jet mode).

In the cosmological zoom simulations that we study here, in addition
to thermal energy, we also deposit momentum associated with
radiatively efficient accretion into the gas particles surrounding the
black hole, and, further, model photo-ionization and photo-heating
by this radiation. The prescriptions for momentum and radiation
feedback from AGN (MrAGN) presented in \citet{Choi2012, Choi2014,
  Choi2015} have been implemented into SPHGAL \citep{Hu2014}, an
updated version of the Smoothed Particle Hydrodynamics code \texttt{GADGET-3}
\citep{Springel2005}. The code has also recently been modified to
include updated treatments of many physical processes, including an
improved treatment of stellar and supernova feedback
\citep{Nunez2017}, chemical enrichment and metal line cooling \citep{Aumer2013},
photoelectric heating, and X-ray heating by the meta-galactic X-ray
background. These updates are described in detail in \citet{Choi2016}.

We address several questions: 1) what are the histories of inflow and
outflow for these galaxies? 2) how much of the material ejected by
winds is removed permanently and how much comes back, and on what
timescales? and 3) how do these winds affect the host galaxy? We
compare simulations including both AGN feedback and feedback due to
stars and supernovae (MrAGN) to a matched set of simulations that
include only stellar and supernova feedback (NoAGN). In this way we
can attempt to isolate the effect of the AGN-driven winds on the gas
cycle in these halos, which have virial masses of
$\sim10^{12}\--10^{13.4}M_{\odot}$, and consider how the answers to
the questions posed above differ in the MrAGN vs. NoAGN cases.

In Section 2, we describe the simulations used as well as our analysis
methods and in Section 3 we present our results. In Section 4 we
discuss our results. We summarize in Section 5. 

\section{Simulation and Methods}

In this work we present simulated massive galaxies from cosmological
zoom-in runs performed with a version of the parallel SPH code
GAGDET-3 \citep{Springel2005}. Specifically, we use the modified code,
SPHGAL which mitigates problems previous SPH codes have traditionally
had with fluid mixing \citep{Hu2014}. A detailed description of the
code can be found in \citet{Hu2014} and relevant updates can be found
in \citet{Choi2016}, but below we give a brief overview of the
physics relevant to our study.

\subsection{Code Basics and Setup}

The code employs the pressure-entropy SPH formulation of
\citet{Hopkins2013} and also has improved force accuracy due to the
use of the Wendland $C^{4}$ kernel with 200 neighboring particles
\citep{Dehnen2012}. We also include the improved artificial viscosity
implementation presented by \citet{Cullen2010} and an artificial
thermal conductivity according to \citet{Read2012} in order to reduce
the noise in pressure estimates in the presence of strong
shocks. Finally, a timestep limiter is employed according to
\citet{Saitoh2009} and \citet{Durier2012} to ensure that neighboring
particles have similar timesteps and that ambient particles do not
remain inactive when a shock is approaching.

\subsection{Star Formation and Stellar/Supernova Feedback}

Star formation and chemical evolution are modeled as described in
\citet{Aumer2013}; chemical enrichment is achieved via winds driven by
Type I and II supernovae and asymptotic giant branch (AGB)
stars. Eleven species of metals are tracked explicitly and cooling
rates are calculated based on abundances, as well as the temperature
and density of the gas. Redshift-dependent metagalactic UV/X-ray and
cosmic microwave backgrounds are included with a modified
\citet{Haardt2012} spectrum.

Stars are formed stochastically where the density is greater than the 
density threshold for star formation. This threshold is given as
$n_{\rm{th}}\equiv
n_{0}(T_{\rm{gas}}/T_{0})^{3}(M_{0}/M_{\rm{gas}})^{2}$ where $n_{0}$ =
2.0 $\rm{cm}^{-3}$ and $T_{0} = 12000$ K, with $M_{0}$ being the gas
particle mass. This corresponds to the density value for the Jeans 
gravitational instability of a mass $M_{\rm{gas}}$ at temperature 
$T_{\rm{gas}}$. The star formation rate is calculated as 
$d\rho_{*}$/$dt$ = $\eta\rho_{\rm{gas}}$/$t_{\rm{dyn}}$, where 
$\rho_{*}$, $\rho_{\rm{gas}}$, and $t_{\rm{dyn}}$ are the stellar 
density, gas density and local dynamical time for the gas particle, 
respectively. $\eta$ is the star formation efficiency and is set to 0.025.

Stellar feedback is included in the form of stellar winds and heating
by ionizing radiation from young massive stars. Momentum from stellar
winds is added to the surrounding gas particles, while cold gas within
the Str{\"o}mgren radius of hot stars is heated to $T=10^{4}$ K.

In the supernova feedback model, SN energy and momentum
is distributed to the surrounding ISM from the SN event. Depending 
on the distance from the SN events, we assume that each nearby 
gas particle is affected by one of three successive phases of SN 
remnant (SNR) evolution: ejecta-dominated free expansion (FE) phase, 
energy-conserving Sedov-Taylor blast-wave SNR phase, and 
momentum-conserving snowplow phase. SN energy is transferred by 
conserving the ejecta momentum for gas particles within the radius of the FE 
phase. For the gas particles lying outside the FE radius but within the 
Sedov-Tayler phase, the supernova energy 
is transferred as 30\% kinetic and 70\% thermal. Finally at larger
radii in the snowplow phase, a fraction of the original SN energy is transferred
as radiative cooling becomes significant. An allowance is also made for 
more efficient propagation of the remnant in a multi-phase interstellar 
medium where appropriate. A detailed description of the implementation 
of the early stellar and supernova feedback prescription can be found 
in \citet{Nunez2017}.

Feedback from low- and intermediate-mass stars is also included in the
form of slow winds as from AGB stars. Energy and momentum are
transferred to surrounding gas particles such that momentum is
conserved. The initial outflowing wind velocities are assumed to be
$v_{\rm{out,AGB}}=10$ km $\rm{s^{-1}}$, which is typical for AGB stars
\citep{Nyman1992}. Metal-enriched gas from all of these prescriptions
is continuously added to the ISM. We also include metal diffusion, which 
allows for the mixing and spreading of metals in the enriched gas 
(see \citet{Aumer2013} for details).

\subsection{Black Hole Growth and Feedback}

Seed black holes are treated as collisionless sink particles and are
placed in the centers of haloes that reach a mass threshold
($10^{5.15} M_{\odot}$ black holes are placed in $10^{11.15}
M_{\odot}$ haloes). These black holes can then grow by gas accretion
or by merging with another black hole, as soon as the two black holes
fall within each other's local SPH smoothing lengths and their
relative velocities are smaller than the local sound speed. In the
case of gas accretion, infall onto the black hole is governed by the
Bondi-Hoyle-Littleton rate \citep{Bondi1952}. The size of the SPH gas
particles is taken into account so that full accretion is allowed when
the entire volume of a particle is within the Bondi radius. Gas
particles which fall only partially inside the Bondi radius are given
a probability of being absorbed by the black hole based on the volume
within the Bondi radius \citep{Springel2005b, Choi2012, Choi2014}.

%rss i think winds can originate from a variety of scales both smaller and
%larger than the accretion disk
Massive disk winds driven from the nuclear region
surrounding the central supermassive black hole are thought to 
produce many observed AGN spectral features, including the broad absorption
lines \citep[e.g.][]{Knigge2008}. These wind-driving processes on scales 
below our resolution limit,
such as radiation pressure acting on spectral lines,  are treated as
sub-grid processes and assumed to impart momentum to the gas.  AGN driven 
winds are launched from the central region around the black hole with a fixed
wind velocity of $v_{\rm outf,AGN}=10,000$ km/s, and the number of 
particles selected to receive a kick due to this wind is determined by a 
parameter for feedback efficiency.  The total energy flux carried by 
the wind is $\dot{E}_{\rm{w}}\equiv\epsilon_{\rm{w}}\dot{M}_{\rm{acc}}c^{2}$, 
where the efficiency parameter $\epsilon_{\rm{w}}$ is set to 0.005 
\citep{Ostriker2010, Choi2016}. The mass  flux and momentum flux carried by
the wind are $\dot{M}_{\rm outf} = 2 \dot{M}_{\rm acc} \epsilon_{\rm{w}} 
c^2 / v_{\rm outf,AGN}^2$ and $\dot{p}_{\rm outf} = 2 \epsilon_{\rm w} 
\dot{M}_{\rm acc} c^2 / v_{\rm outf,AGN}$ 
respectively, and therefore for our selected wind velocity and
feedback efficiency we have $\dot{M}_{\rm outf} = 9 \dot{M}_{\rm acc}$
and $\dot{p}_{\rm outf}= 30 L_{\rm BH}/c$.
Thus, 90\% of the inflowing mass entering the central region is expelled 
while 10\%  is accreted onto the black hole. 

The selected gas particles receive the wind kick in a direction parallel or
anti-parallel to their angular momentum vectors. The emitted wind particle
shares its momentum with its two nearest neighbors to reproduce
the shock heated momentum-driven flows. The momentum is split between the
gas particles and is conserved, but the kinetic energy of the outflow
is not; the excess energy is deposited into the gas particles as
thermal energy. This gives 2:1 divisions into thermal and kinetic energies
which is similar to that in the Sedov-Taylor blast wave, making the shocked
winds approach the Sedov solution faster, thus making us less sensitive to 
the resolution.

Radiation feedback from Compton and photoionization heating due to
X-ray radiation from the accreting black hole, radiation pressure
associated with the heating, and the Eddington force are also
included. We utilize the AGN spectrum and metal line heating
prescription of \citet{Sazonov2004}. X-ray radiation is coupled to
surrounding gas using an approximation from \citet{Sazonov2005}. The
radiation pressure on each gas element is also calculated. Accretion
is not artificially capped at the Eddington rate, but the Eddington
force acting on electrons is included such that super-Eddington
accretion can occasionally occur but naturally reduces inflow while
stimulating outflows.

Also included are metallicity-dependent heating prescriptions due to
photoelectric emission and metal line
absorption. The emissivity of background AGN is calculated as 
$\varepsilon(z)=\epsilon\frac{d\rho_{\rm{BH}}(z)}{dt}c^{2}$, where 
the radiative efficiency $\epsilon$ is set to 0.1. From this the heating 
by the cosmic X-ray background is derived. More details about these 
and all of our feedback prescriptions can be found in \citet{Choi2016}.

We note here that these feedback prescriptions have been shown to
produce galaxies with fairly realistic observable properties
\citep{Choi2016}. Without AGN feedback, our model produces galaxies
with stellar to halo mass ratios which are too high by a factor of
three, and continue to form stars until $z=0$, in conflict with
observations of galaxies of this mass, which are predominantly
quenched. This results in compact stellar cores, effective radii which
fall a factor of five below the observed size-mass relation and high
velocity dispersions \citep{Choi2016}. The addition of AGN feedback
alleviates all of these problems, resulting in quiescent galaxies with
observationally consistent sizes and velocity dispersions at $z=0$. Although
AGN feedback alleviates the over-cooling problem and results in galaxy
sizes in good agreement with observations, our lower mass galaxies, and 
high-redshift galaxies, tend to have sizes that are too small for their 
mass compared with observations \citep{Choi2016}. While the cold gas 
fractions of our galaxies are lower than observed at low redshift due to 
shortcomings in our treatment of the ISM and stellar feedback,
AGN feedback effectively removes hot gas from halos, resulting in lower, more
physical X-ray luminosities and total gas mass fractions
\citep{Choi2016}. Post-processing the MrAGN simulations with newly 
developed nebular emission line models additionally shows that the 
evolution of optical nebular emission line-ratios of massive galaxies is 
widely consistent with observations \citep{Hirschmann2017}. Satisfied 
that our model produces galaxies which compare
favorably with observed high-mass elliptical galaxies, in this work we
turn to analyze the wind properties and gas cycle in these systems.

\subsection{Zoom Simulations}

The ``zoom-in'' initial conditions that we use are described in detail
in \citet{Oser2010} and \citet{Oser2012}. We run our zooms in dark
matter haloes picked from a dark matter only simulation which employed
the parameters from WMAP3 \citep{Spergel2007} and assumed a flat
cosmology: $h=0.72, \Omega_{\rm{b}}=0.044, \Omega_{\rm{dm}}=0.216,
\Omega_{\rm{\Lambda}}=0.74, \sigma_{\rm{8}}=0.77$, and an initial
power spectrum slope $n_{rm{s}}=0.95$. We trace back the dark matter
particles close to the haloes of interest in each snapshot and then
replace those dark matter particles with high-resolution dark matter
and gas particles. The high-resolution zoom is then evolved from $z=43$
to today.

The selected dark matter haloes have final virial masses between
1.4$\times 10^{12}M_{\odot}$ and 2.3$\times 10^{13}M_{\odot}$ and are
made up of dark matter particles with a mass of
$m_{\rm{dm}}$=3.57$\times 10^{7}M_{\odot}$. The final central galaxy
masses in these halos (for our runs with AGN) are between 8.2$\times
10^{10}M_{\odot}$ and 1.5$\times 10^{12}M_{\odot}$, with gas and star
particles both having a mass of $m_{*,\rm{gas}}$=6.0$\times
10^{6}M_{\odot}$. We use comoving gravitational softening lengths of
$\epsilon_{\rm{gas, star}}$=.556 kpc and $\epsilon_{\rm{halo}}$=1.236
kpc for gas/star particles and dark matter particles, respectively.

In what follows, we will examine zoom regions run with two different
models: MrAGN and NoAGN. MrAGN is the fiducial model which includes
all of the physics described above, including the different ways in
which AGN can effect feedback on their surrounding galaxies. NoAGN is
the same in every way except it contains no black holes and thus no
AGN feedback. In this way the effects of the different feedback
mechanisms can be isolated. Out of our initial sample of 30 pairs of
zoom runs, we focus on the 24 of which we can use to make a direct
comparison between the two models, as described below.

\subsection{Methods}

\subsubsection{Galaxy Matching}

In order to compare MrAGN galaxies with their NoAGN counterparts, and
also to examine inflowing and outflowing gas in our simulations, we
must first find the center of our haloes of interest and make sure
that we are tracking the same progenitor back with redshift in both
runs. We utilize two different codes to track the halo centers:
ROCKSTAR \citep{Behroozi2013} and GTRACE, a tool built specifically
for analyzing GADGET snapshots. In 24 of our 30 pairs, at least one of
these tools found the center of the same progenitor in both runs by
$z=1$. Going back to $z=2$, we have 23 matched pairs and by $z=3$ we
have 21 matched pairs. There are brief periods when the center even of
these matched pairs is lost, such as when a satellite of comparable
mass approaches the main progenitor, but this is generally only for a
short time and can be easily identified when a relatively stable
quantity, such as stellar mass, suddenly dips as can be seen in Figure
\ref{flowplot_163} below. We exclude the six galaxies for which the found 
centers do not correspond to the same progenitors by $z=1$. 

\subsubsection{Particle Tracking}

With the centers found, we track the flow of particles across two
shells: one at 10\% of the virial radius and one at the virial
radius. We refer to the former as the ``galaxy radius'' $\rgal$ and the
latter as the ``halo radius'' $\rhalo$.  The tracking begins when there is 
a reliable center found for both the MrAGN and NoAGN runs; for our 
three case study galaxies, for which it is important to track over the 
same period (as we display cumulative quantities), this tracking begins 
at z $\sim$ 4, give or take one timestep. Almost all of the rest of our 
galaxies are tracked by z=3, as mentioned above. At each timestep the gas
particles which are inside both of these radii are catalogued. In the
next timestep any particle that was inside (outside) the radius of
interest in the last timestep and is now outside (inside) that radius
and has a positive (negative) radial velocity is considered outflowing
(inflowing). Note that the typical time step between snapshots at 
around z=0 is 0.13 Gyr.

The inflowing and outflowing mass at each timestep is
stored, and the inflow and outflow rate can be calculated by using the
timestep between snapshots. We also keep track of whether a gas
particle is accreting or outflowing for the first time, or if it has
done so previously. Finally, we keep track of how long it takes for
individual gas particles to be recycled, as well as the maximum radial
displacement experienced by the particle during each recycling event
(see \citet{Uebler2014} for a similar treatment of gas particles in
disc galaxies). Unless otherwise specified, stellar and gas masses of the 
galaxy are calculated within 10\% of the virial radius.

\section{Results}
\begin{table*}
  \begin{center}
  \caption{Final properties of three example galaxies}
     \vskip+0.1truecm	
{
  \begin{tabular}{c|c|c|c|c|c|c|c|c|c}\hline\hline
    Galaxy & Halo Mass & Stellar Mass & BH Mass & Hot Gas Mass & Inflow$_{\rm{g}}$ & Outflow$_{\rm{g}}$ & Inflow$_{\rm{h}}$ & Outflow$_{\rm{h}}$ & $M_{\rm{in,h}}$/$f_{\rm{b}}M_{\rm{h,0}}$ \\
    \hline\hline
    m0163 (MrAGN) & 13.1 & 11.4 & 9.2 & 11.9 & 11.2 & 11.2 & 12.2 & 11.9 & 0.90 \cr
    \hline
    m0163 (NoAGN) & 13.1 & 11.8 & N/A & 12.1 & 11.8 & 11.5 & 12.3 & 11.6 & 1.00 \cr
    \hline
    m0329 (MrAGN) & 12.7 & 11.3 & 9.2 & 11.2 & 10.9 & 10.7 & 11.9 & 11.7 & 0.85 \cr
    \hline
    m0329 (NoAGN) & 12.8 & 11.7 & N/A & 11.7 & 11.5 & 10.9 & 12.0 & 11.1 & 1.05 \cr
    \hline
    m0501 (MrAGN) & 12.5 & 11.2 & 9.3 & 7.3 & 11.0 & 10.8 & 11.5 & 11.4 & 0.57 \cr
    \hline
    m0501 (NoAGN) & 12.6 & 11.5 & N/A & 11.4 & 11.4 & 10.8 & 11.8 & 10.9 & 1.00 \cr
    \hline\hline
  \end{tabular}}
  \end{center}
  {\bf Note.} The final halo, stellar, black hole and hot gas masses, the 
    cumulative gas inflow and outflow masses that crossed
    shells at the galaxy radius and the halo radius, and the cumulative 
    inflow mass at the halo radius divided by the final halo mass times 
    the universal baryon fraction for our three example galaxies,
    both for our MrAGN run and our NoAGN run. All masses are given in
    units of log solar masses.
\end{table*}

Here we present the results of our analysis, first examining the
detailed histories of a few representative galaxies, then looking at
broader trends in galaxy and inflow/outflow properties for the entire
set of galaxies. We note again that outflows are driven by both stars
and supernovae and AGN. Our analysis also captures inflow and outflow
due simply to thermal motions of gas. This is more apparent in our
NoAGN galaxies, as will be mentioned below.

\subsection{Case Studies}

When examining the histories of individual galaxies in terms of their
basic properties and the properties of their inflows and outflows,
some broad classes emerge, mainly as a result of galaxy mass. Below we
present three galaxies representative of their mass bins: ``high
mass'' m0163 ($M_{\rm{*,final}}\sim10^{11.4}M_{\odot}$ and
$M_{\rm{h,final}}\sim10^{13.1}M_{\odot}$), ``intermediate mass'' m0329
($M_{\rm{*,final}}\sim10^{11.3}M_{\odot}$ and
$M_{\rm{h,final}}\sim10^{12.7}M_{\odot}$), and ``low mass'' m0501
($M_{\rm{*,final}}\sim10^{11.2}M_{\odot}$ and
$M_{\rm{h,final}}\sim10^{12.5}M_{\odot}$). We study the baryon cycles
in these halos in detail, and use them as exemplars of more general
trends that we discuss later.

\subsubsection{Galaxy Property and Gas Flow Histories}

\begin{figure*}
\centering
  \includegraphics[width=0.6\textwidth]{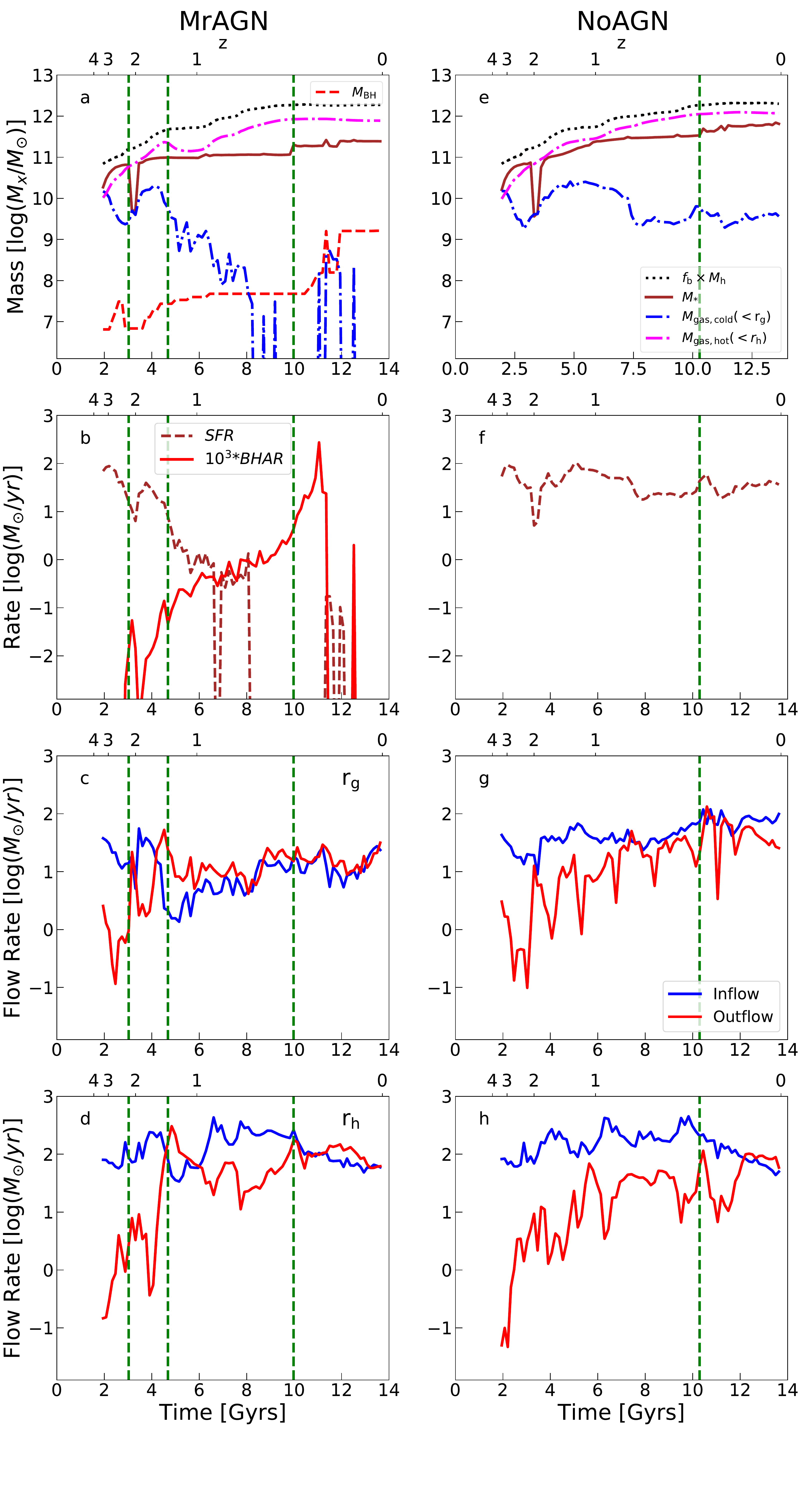}
\caption[History of galaxy m0163]{Baryon cycle history of the main progenitor 
  galaxy in halo m0163. Left Column: Results for MrAGN run. Right Column: 
  Results for NoAGN run. Top row: Evolution of galaxy properties -- Baryon
  fraction times halo mass, stellar mass, total black hole mass (for MrAGN
  run), cold gas mass within the galaxy radius (T<2$\times10^{4}$K) and hot gas
  mass within the halo radius (T>2$\times10^{4}$K). {\bf Note that we show total black
  hole mass within the galaxy radius $\rgal$.} Second row: Evolution of 
  star formation rate and black hole accretion rate (for MrAGN run; black hole 
  accretion rate is scaled up by a factor of
  1000). Third row: Inflow and outflow rates at the galaxy radius. Fourth row:
  Inflow and outflow rates at the halo radius. Green vertical dashed lines indicate
  halo merger events for which the halo mass ratio is 1:10 or
  greater. The final stellar mass in the MrAGN run is smaller due to
  the decrease in cold gas supply and thus star formation rate. AGN feedback does not
  affect the inflow of gas into the halo, but does mildly affect the
  inflow rate of gas at galactic scales. It also enhances outflows on
  both scales at early times.}  {\label{flowplot_163}}
\end{figure*}

Halo m0163 has a history characteristic of the more massive end of our
population. It has a final halo mass of log($M_{\rm{h}}/M_{\odot}$)
$\sim$ 13.1 and a final stellar mass of log($M_{*}/M_{\odot}$) $\sim$
11.4 in the MrAGN run. Figure \ref{flowplot_163} shows the
evolutionary histories of different galaxy, black hole, inflow and
outflow properties. In panels (a) and (e) we see that the halo mass
(shown multiplied by the universal baryon fraction, 0.1658) is largely
unaffected by AGN feedback, while the final stellar mass is reduced by
roughly 0.4 dex. While the cold gas mass (T<2$\times10^{4}$ K)
decreases only mildly in the NoAGN run, all of the cold gas within
$r_g$ is removed or heated by AGN feedback in the MrAGN run. The hot
gas mass of the MrAGN galaxy is only slightly smaller than that for
the NoAGN galaxy, due to removal of some of the hot gas from the
halo. We can also see the growth of the black hole in the MrAGN run,
which reaches a final mass of $\log(M_{*}/M_{\odot})=9.2$. The green
dashed lines denote mergers for which the halo ratio is greater than
1:10.

Panels (b) and (f) show that as the cold gas mass decreases, the star
formation rate falls just as drastically, which is partially
responsible for the smaller final stellar mass found in the MrAGN
run. Star formation continues at a nearly constant rate in the NoAGN
run (several tens of solar masses per year). We also see the black
hole accretion rate (BHAR), which gradually increases from $\sim$ 4
Gyrs until about 12 Gyr after the start of the simulation. After this
point, black hole accretion itself is quenched.

Panels (c), (d), (g) and (h) show the inflow and outflow rates for
both runs on galaxy scales $\rgal$ and halo scales $\rhalo$. In the
MrAGN run, the inflow rate at $\rgal$ is suppressed relative to
the NoAGN run value of $\sim$ 50 $M_{\odot}$/yr to only 1-2
$M_{\odot}$/yr, and overcome by the outflow rate after the two early
merger events at $\sim$ 3-5 Gyrs (just after $z\sim$ 2). The outflow
rate continues to mirror the inflow rate thereafter at $\sim$ 10-30
$M_{\odot}$/yr. At $\rhalo$, this same spike in outflow rate occurs
after a slight delay, due to material that was caught up in the
initial outflow at $\rgal$ and eventually crosses the virial radius,
pushing out even more material on its way. This results in outflow
rates of as much as $\sim$ 300$M_{\odot}$/yr. The outflow rate at
$\rhalo$, however, rarely overtakes the inflow rate, even in the MrAGN
run. In the NoAGN run, inflow dominates at almost all times at both
radii. The outflow rate at $\rgal$ in the NoAGN case steadily rises,
eventually reaching an equilibrium with the inflow rate at a value
higher than is seen in the MrAGN case. This is due to the fact that we
have more gas in the central region in the NoAGN case which is cycling
in and out of the galaxy due to thermal motions. The inflow rate at
$\rhalo$ is largely unaffected by AGN feedback, while the outflow rate
in the MrAGN run is enhanced relative to the NoAGN case. This means
that the AGN feedback acts mostly in an ejective way at this halo mass
($\sim10^{13.1}M_{\odot}$) at halo scales, rather than
preventatively. See Table 1 for the cumulative inflow and outflow gas
masses that cross both shells over the duration of the
simulation. Already apparent from Table 1 is the importance of
preventative feedback, as represented by the cumulative mass in
baryons that accreted onto the halo divided by the final halo mass
times the universal baryon fraction.

\begin{figure*}
\centering
  \includegraphics[width=0.6\textwidth]{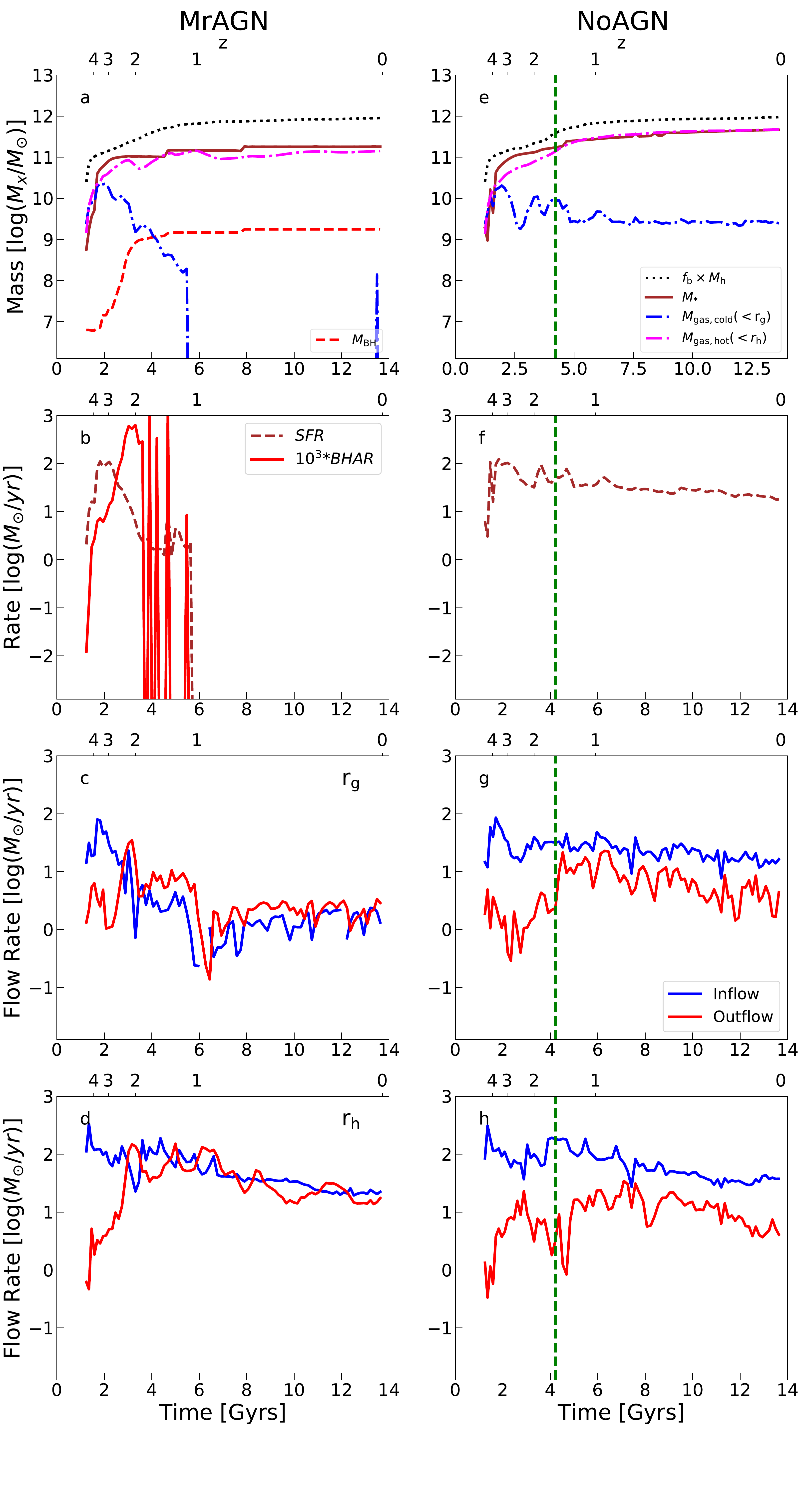} 
\caption[History of galaxy m0329]{Same as Figure \ref{flowplot_163}, but for halo m0329. Inflow
  in the MrAGN case is more noticeably suppressed at late times at both
  radii of interest than in the more massive halo m0163. Following an
  initial outflow event at $\sim4-6$ Gyrs, outflow dominates inflow at
  the galaxy radius by a small amount, but consistently. In the NoAGN case
  inflow dominates outflow at all times at both radii.}
        {\label{flowplot_329}}
\end{figure*}

Figure \ref{flowplot_329} is the same as Figure \ref{flowplot_163},
but now for galaxy m0329. This galaxy falls near the middle of our
mass range, with a final halo mass of log($M_{\rm{h}}/M_{\odot}$)
$\sim$ 12.7 and a final stellar mass of log($M_{*}/M_{\odot}$) $\sim$
11.3 in the MrAGN run. We see again the difference between the stellar
mass (again $\sim$ 0.4 dex) and cold gas mass in the MrAGN run versus
those in the NoAGN run. The galaxy in the MrAGN run has a steeply
decreasing cold gas mass, resulting in a steeply decreasing SFR. In
the case of m0329, this also corresponds to an increasing black hole
accretion rate. The suppression of inflow at both radii of interest is
more pronounced than for m0163; the inflow rate at $\rgal$ is
significantly (and permanently) decreased from >10 $M_{\odot}$/yr to
1-2 $M_{\odot}$/yr following a strong outflow event at $\sim4\--6$
Gyrs, concluding at $z\sim1$. The inflow rate at $\rhalo$ is also more
noticeably decreased after this outflow event than it was for the more
massive halo m0163; the final inflow rate at $\rhalo$ is $\sim$ 1.7
times larger in the NoAGN run than in the MrAGN run. This suggests
that as we look at smaller halo masses and therefore shallower
potential wells, \textit{preventative feedback}, where AGN feedback
not only removes material but also prevents new material from
accreting, becomes more important (see the rightmost column of Table
1). The outflow rates at both radii match or exceed the inflow rates
at almost all times after the initial outflow event in the MrAGN
run. In the NoAGN case, inflows and outflows behave much as they did
for m0163, with inflow always dominating.

\begin{figure*}
\centering
  \includegraphics[width=0.6\textwidth]{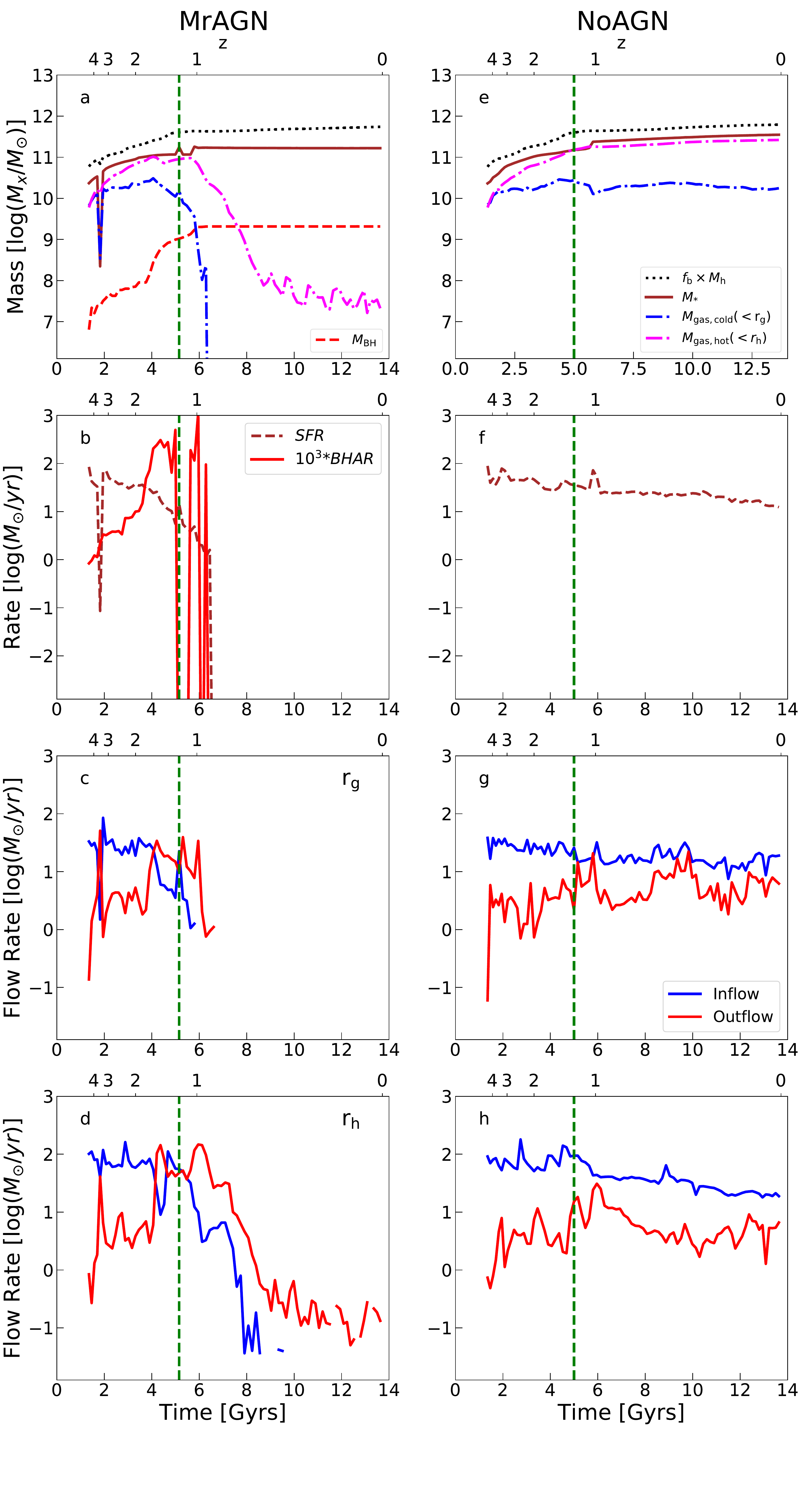}
  
\caption[History of galaxy m0501]{Same as Figures \ref{flowplot_163} and \ref{flowplot_329},
  but for halo m0501. While the NoAGN case is similar to the previous
  galaxies, the MrAGN case is exceptional in that AGN feedback
  completely clears the galaxy of gas and prevents any new gas from
  accreting for several billion years, even on the scale of the halo.}
        {\label{flowplot_501}}
\end{figure*}

Finally, Figure \ref{flowplot_501} is the same as Figures
\ref{flowplot_163} and \ref{flowplot_329}, but now for galaxy
m0501. This galaxy is less massive than m0329 with a final halo mass
of log($M_{\rm{h}}/M_{\odot}$) $\sim$ 12.52 and a final stellar mass
of log($M_{*}/M_{\odot}$) $\sim$ 11.22 in the MrAGN run, and has a
history more characteristic of the low mass galaxies in our suite. In
the top four panels, the galaxy properties of m0501 evolve very
similarly to those already examined for m0163 and m0329, except for
hot gas in the MrAGN run, which is much more strongly affected. When
focusing on the inflow and outflow properties, m0501 is very
different. In the NoAGN case, things are much the same, with inflow
dominating outflow at both radii at almost all times. In the MrAGN
case, several bursts of outflow of between 10-30 $M_{\odot}$/yr
dominate inflow at $\rgal$ at around 4 Gyrs. This outflow is powerful
enough to halt inflow, after which the outflow stops as well because
the gas within $\rgal$ has been completely depleted. At $\rhalo$, we
see this outflow, having swept up gas in the halo and now removing
$\sim$ 100 $M_{\odot}$/yr, peak at a slightly later time and once
again halt inflow across the virial radius. This outflow then also
tapers off as it has cleared most of the galaxy's halo gas as
well. This is an extreme case where AGN feedback acts in both an
ejective and a preventative way on dark matter halo scales. This is
fairly rare, occurring only in 4 of our 30 MrAGN galaxies. These
galaxies all have relatively small halo masses for our sample (the
most massive has $M_{\rm{h}}\sim10^{12.5}M_{\odot}$) and all contain a
relatively large black hole mass for their halo mass, although they
still sit on the $M-\sigma$ relation.

\subsubsection{Gas Morphology and Velocity Structure}
\begin{figure*}
  \includegraphics[width=1.0\textwidth]{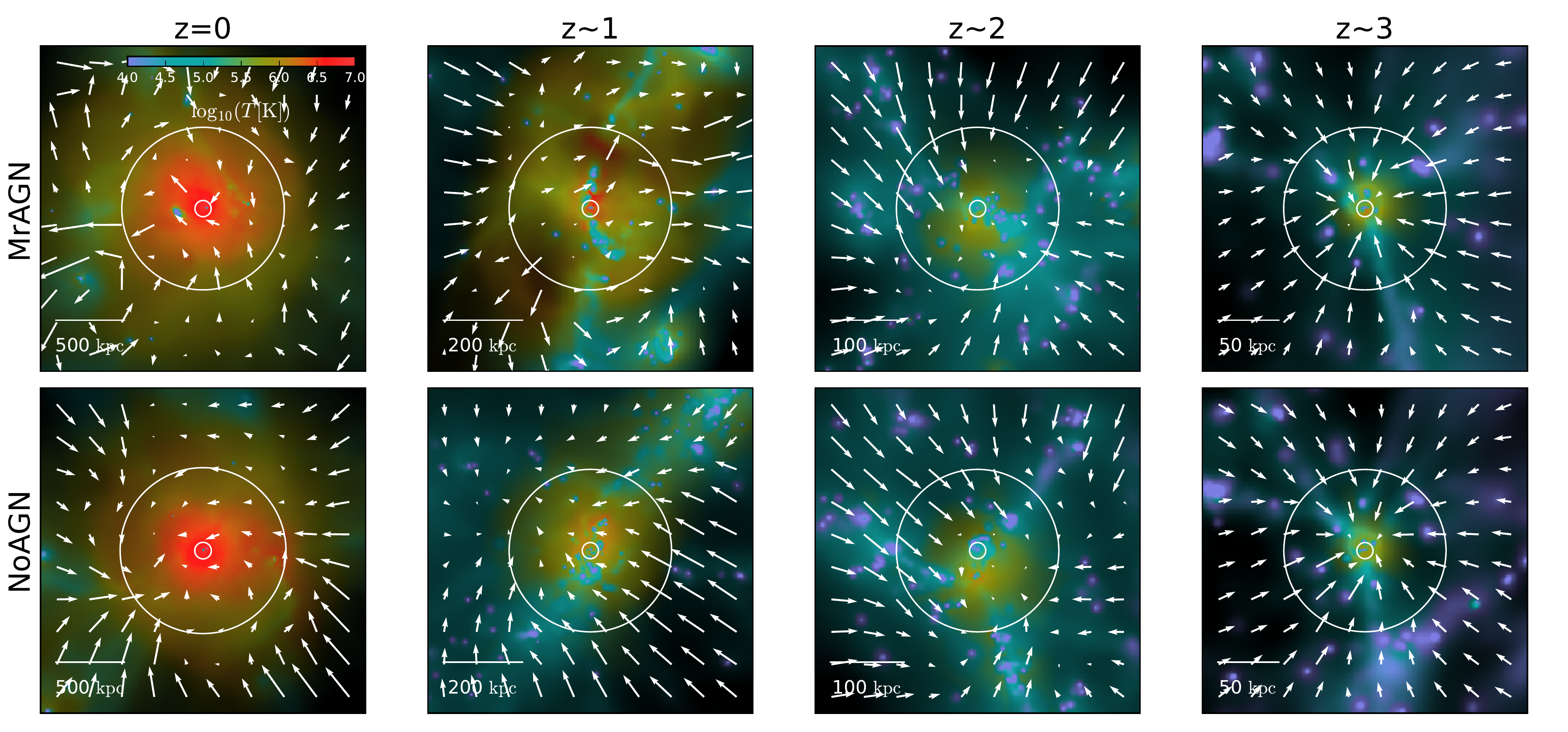}
  
  \caption[Velocity map of m0163]{Gas temperature and velocity vector maps for galaxy
    m0163. Images created using \texttt{pygad}
    \citep{Roettgers2017}. In the gas image, the color corresponds to
    gas temperature, while the luminance corresponds to gas
    density. The velocity vectors are calculated for a slice with a
    thickness of the halo radius at the redshift of interest, with
    contributing gas velocities weighted by their densities. The arrow
    lengths are normalized by the number of bins, and then normalized
    by the average arrow length. Top row: Results for MrAGN
    run. Bottom row: Results for NoAGN run. The velocity vectors are
    overlaid onto gas temperature maps at four redshifts. White
    circles denote the current the galaxy radius and the halo radius. At
    $z\sim3$, the two runs are nearly identical and dominated by
    accretion along filaments. At $z\sim2$, the two runs are still
    very similar, with the bulk velocity flow due to gas inflowing onto
    the central galaxy. At $z\sim1$, the MrAGN run is undergoing a
    bout of outflow, while the NoAGN run is still steadily accreting
    gas. At $z\sim0$, though the remnants appear similar, the gas in
    the MrAGN run is more diffuse.}  {\label{vel_vec_163}}
\end{figure*}

In Figure \ref{vel_vec_163} we show vector maps of the gas velocities
in m0163 overlaid onto color maps of the gas temperature for both runs
at four different redshifts. The luminance of the gas in the images
corresponds to its density. At $z\sim 3$ and $z \sim2$, the images of
the two runs look very similar to each other, with the main velocity
features being filaments of gas funneling into the forming galaxy. At
z $\sim1$, however, the two runs look very different. The galaxy in
the MrAGN run is going through a large bout of outflow, which can be
seen in panels (c) and (d) of Figure \ref{flowplot_163} at $\sim5$
Gyrs. Meanwhile, in the NoAGN run, the bulk of the velocity features
are still due to inflowing material onto the galaxy. At $z\sim0$,
while the two runs look very similar, the MrAGN run exhibits a more
diffuse hot gas halo. While gas has been heated by stellar and
supernova feedback, the NoAGN run still exhibits somewhat ordered
motions and inflows as opposed to the MrAGN run, in which we see
material being pushed out of and away from the galaxy.

\begin{figure*}
  \includegraphics[width=1.0\textwidth]{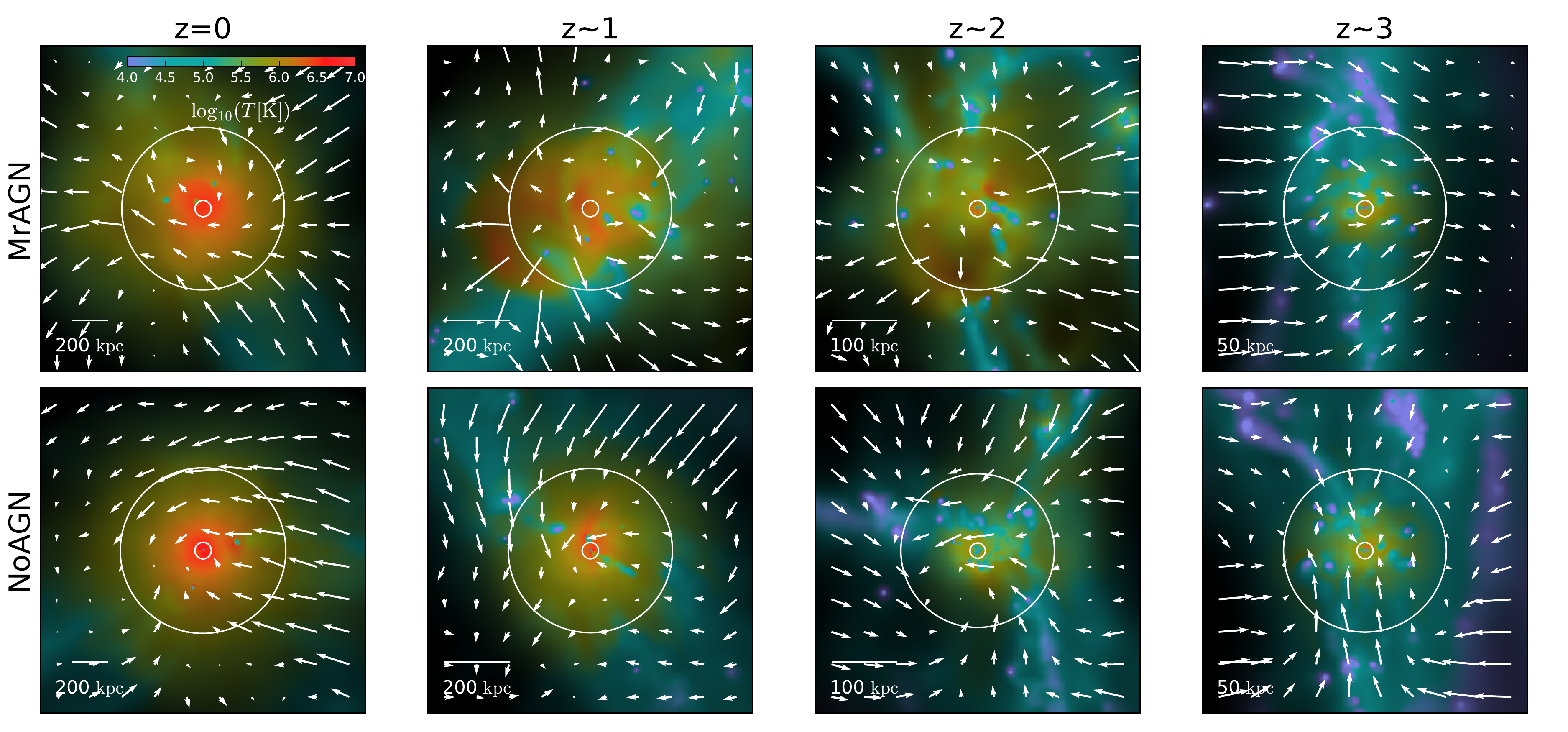}
  
  \caption[Velocity map of m0329]{Same as Figure \ref{vel_vec_163}, but for 
  halo m0329. At $z\sim3$ and $z\sim0$ the two galaxies look very similar, 
  while at $z\sim2$ and $z\sim1$ we see the MrAGN galaxy at the height of 
  its AGN-driven outflow activity in stark contrast to the NoAGN case.}
          {\label{vel_vec_329}}
\end{figure*}

In Figure \ref{vel_vec_329}, the galaxy m0329 in the MrAGN run is undergoing
a major outflow by $z\sim 2$, in contrast to its NoAGN counterpart, which
is dominated by filamentary accretion. At $z \sim1$, while the NoAGN
galaxy also appears to be undergoing feedback as demonstrated by the
bulk of hot gas, there is no strong outflow signature as we see again
in the MrAGN case. Even in the MrAGN run, however, there is still a
strong filamentary inflow feature outside of the outflow sphere of
influence. When this inflowing gas reaches the outflow, some of it is
heated and/or turned around. At $z \sim 0$ we again end up with two
galaxies with similar (to the eye) gas contents, although we know from
Figure \ref{flowplot_329} that there is more to the story.

\begin{figure*}
  \includegraphics[width=1.0\textwidth]{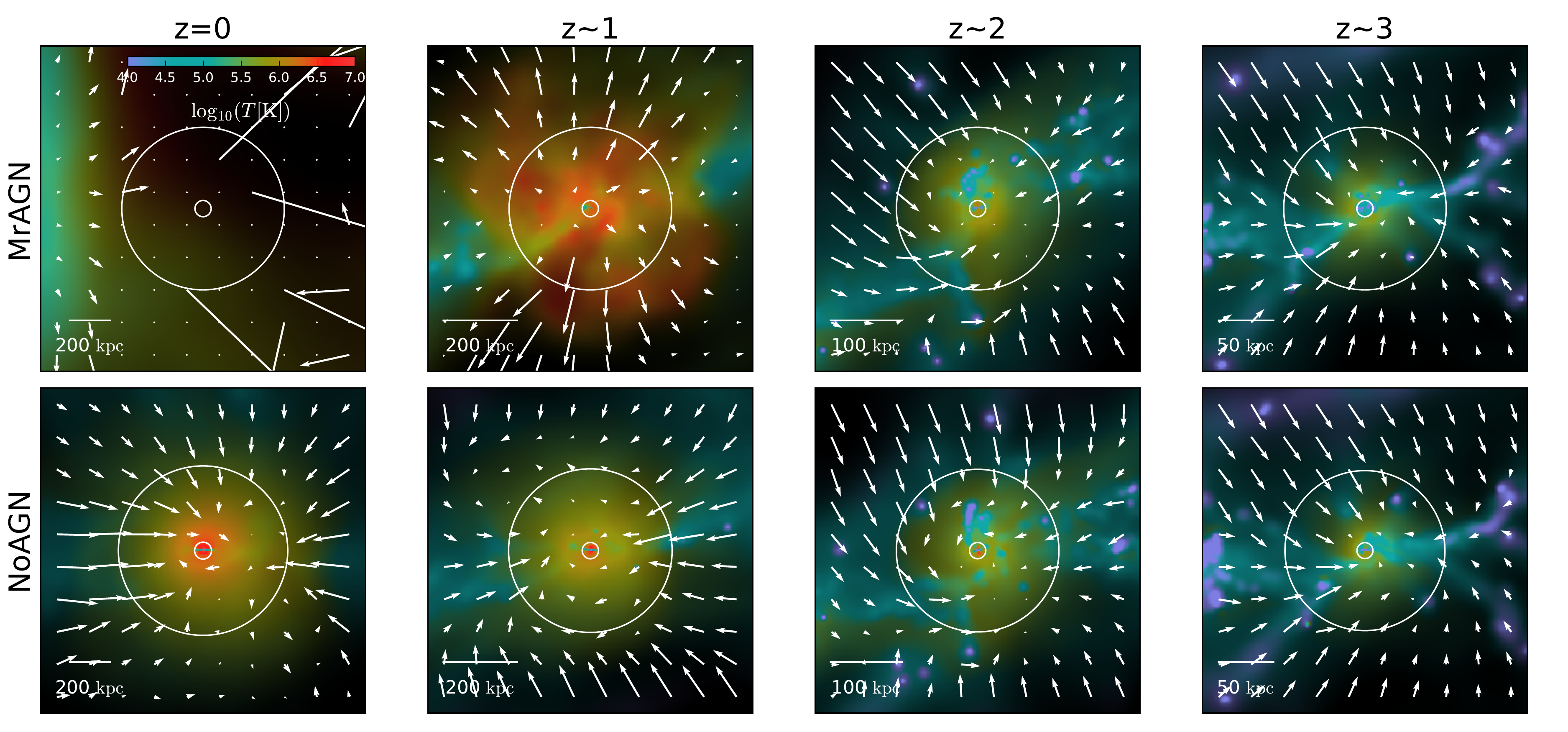}
  
  \caption[Velocity map of m0501]{Same as Figures \ref{vel_vec_163} and 
  \ref{vel_vec_329} but for halo m0501. Between $z\sim1$ and $z\sim0$, 
  AGN-driven outflows drive the gas completely out of the galaxy in the 
  MrAGN run, even destroying the gas filament which was supplying the 
  galaxy with new gas.}
          {\label{vel_vec_501}}
\end{figure*}

Figure \ref{vel_vec_501} shows that the two runs of m0501 are very similar
until $z\sim1$, at which point a large AGN-driven outflow occurs in
the MrAGN case. This outflow goes on to clear the galaxy of gas and
eventually destroy the filament supplying the galaxy with gas, halting
accretion even at the virial radius. At $z\sim0$ there is
basically no gas left to track in the MrAGN run, whereas in the NoAGN
run, the gas has actually settled into a cold disc on galactic
scales. Our mechanical AGN feedback is very efficient at removing gas
from the lower mass galaxies of our sample.

\subsubsection{Recycling Fractions}

\begin{figure}
\centering
  \includegraphics[width=\columnwidth]{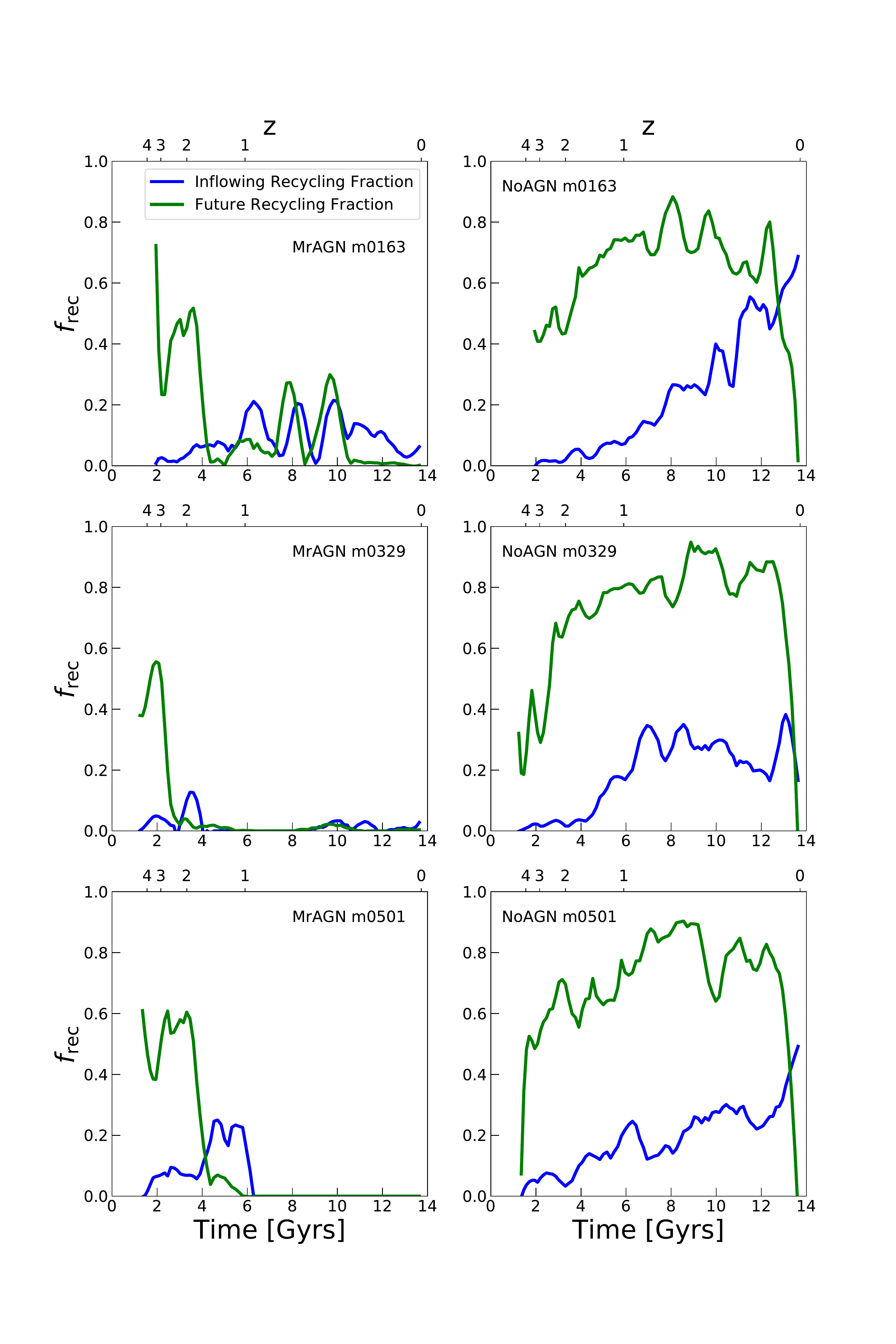}
  
  \caption[Recycling fractions for case studies]{Two different measures of 
  instantaneous recycling fraction for our case study galaxies. The 
  green curves show the future recycling fraction, i.e., the fraction of 
  ejected material that will return at some future time. The blue curves 
  show the inflowing recycling fraction, i.e., the fraction of currently 
  inflowing material that was previously inside the galaxy. Left panel: 
  MrAGN. Right panel: NoAGN. Top panels: m0163. Middle panels: 
  m0329. Bottom panels: m0501. In all cases, both recycling fractions 
  are much higher in the NoAGN runs.}
          {\label{domination_combined}}
\end{figure}

In Figure \ref{domination_combined}, two different measures of the 
instantaneous recycling fraction are plotted. The blue curves represent 
the fraction of inflowing material across $\rgal$ that was previously 
within the galaxy, while the green curves represent {\bf the future 
recycling fraction, i.e., the fraction of currently outflowing material 
across $\rgal$ that will return to the galaxy at some future time. 
The recycling fraction necessarily falls towards zero at late times, 
as outflowing material runs out of time to fall back in before the end 
of the simulation.} From top to bottom we see both the MrAGN and 
NoAGN runs for m0163, m0329, and m0501. In m0163, much of the 
outflowing material at early times in the MrAGN run (50-70\%) is 
destined to come back until the first big bout of outflow at $\sim$ 4 
Gyrs, after which there are only a few isolated incidents of outflowing 
material that will eventually fall back in. This decrease in the future 
recycling fraction also corresponds with the black hole accretion rate 
starting to increase, as can be seen in panel (c) of Figure \ref{flowplot_163}.

As a result, the fraction of inflowing material that is recycled never 
rises much above $\sim$ 20\%. The future recycling fraction in the 
NoAGN run remains very large, between 60 and 90\% for a large 
portion of the galaxy's history. As a result, the inflowing recycling 
fraction by the end of the simulation is almost 70\%. In the MrAGN 
run, of the total $10^{11.2}M_{\odot}$ of gas that accreted onto the 
galaxy throughout its history, $10^{10.1}M_{\odot}$, or $\sim$ 8\% 
was contributed by recycled material. By comparison, in the NoAGN 
run, 31\% of the total $10^{11.8}M_{\odot}$ was recycled material. 
Conversely, the fraction of the cumulative outflow mass that was 
destined to come back when it left is 8.1\% for the MrAGN run and 
67.8\% for the NoAGN run.

The galaxy m0329 behaves similarly to m0163, but there are some 
differences. The drop in future recycling fraction is more extreme 
than in m0163, with the inflowing recyling fraction rarely rising above 
a few percent. In the NoAGN galaxy, the future recycling fraction still 
reaches heights of 90\% and greater, but the inflowing recycling 
fraction doesn't exceed 40\%. This is partially because less material is 
being recycled than in m0163, but also because outflowing material 
falls back in on shorter timescales, leading to a more consistent 
inflowing recycling fraction. The total contribution of recycled material 
to the cumulative inflow mass in the MrAGN galaxy is $\sim$3\%; in 
the NoAGN galaxy it is $\sim$16\%. The fraction of the cumulative 
outflow mass contributing to the future recycling fraction is 5.1\% for 
the MrAGN run and 78.0\% for the NoAGN run.

The galaxy m0501 shows similar trends, except that here there is no 
inflow or outflow in the MrAGN case after $\sim$ 6 Gyrs. We can still 
see that the inflowing recycling fraction by $z=0$ in the NoAGN case 
is $\sim$50\%. The final total contribution of recycled material to the 
cumulative inflow mass is $\sim$9\% for the MrAGN galaxy and 
$\sim$17\% for the NoAGN galaxy. The fraction of cumulative outflow 
mass contributing to the future recycling fraction is 14.2\% in the 
MrAGN run and 69.3\% in the NoAGN run.

\subsubsection{Accretion of Gas and Stars}

\begin{figure}
  \includegraphics[width=\columnwidth]{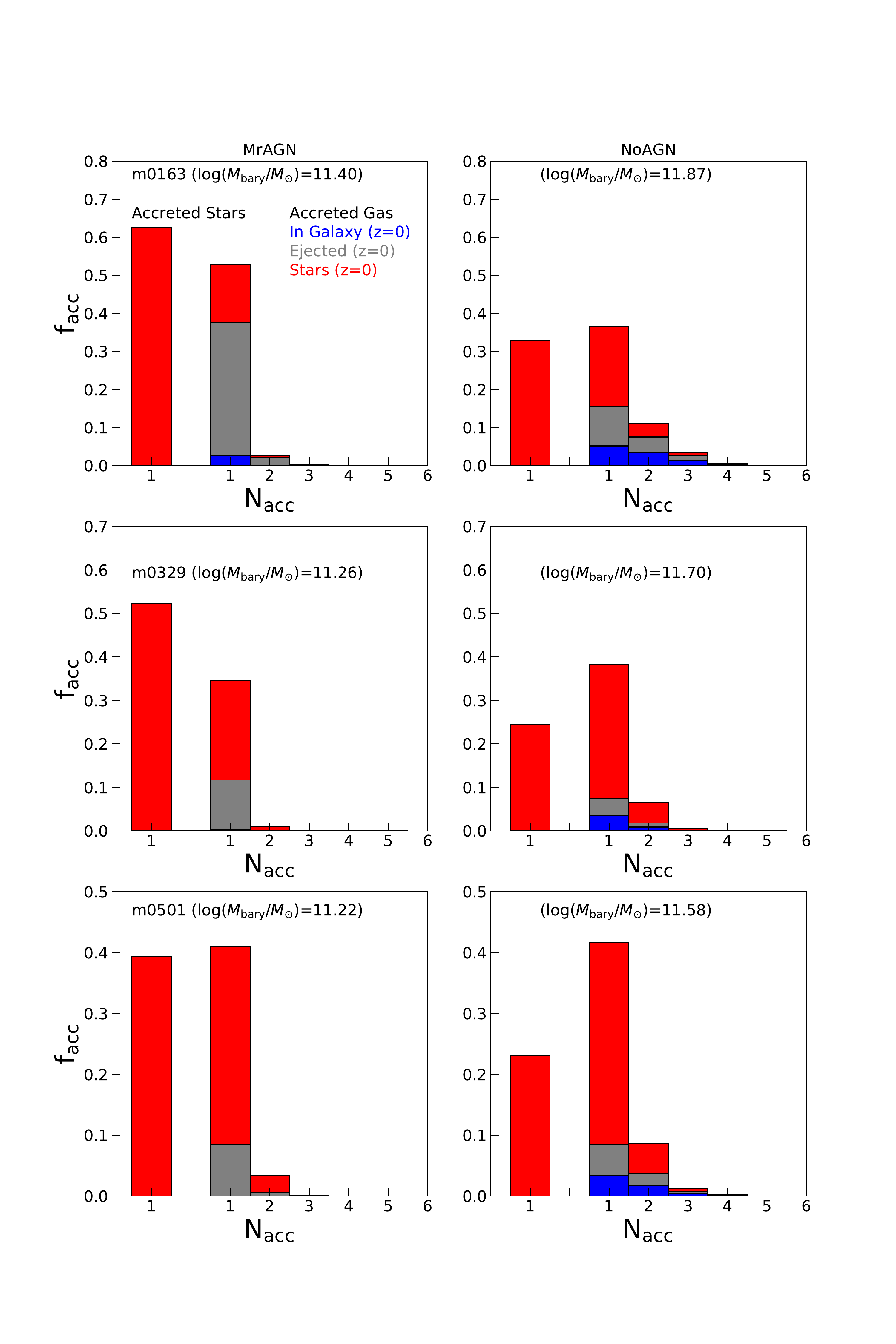}
  
\caption[Accreted stellar fraction and gas recycling events for case studies]
 {The total mass of accreted stars and gas which accreted onto
  the galaxy N times divided by the total baryonic mass of the galaxy
  at $z=0$.  The baryonic mass is given in each panel. Left panels: MrAGN. 
  Right panels: NoAGN. Top panels:   m0163. Middle panels: m0329. Bottom 
  panels: m0501. The far left bar   indicates the mass of accreted stars, 
  which only ever accrete once, divided by the final total baryonic mass 
  within the galaxy radius. The rest of the bars
  indicate gas particles which accreted N times. These bars are
  subdivided into the gas particle's fate at $z=0$: blue if it is
  still gas within the galaxy, grey if it is gas outside of the galaxy
  and red if it is now a star particle. Accreted stars make up a
  larger fraction of the final baryonic mass of MrAGN galaxies than of
  NoAGN galaxies. Gas in the NoAGN run is more likely to accrete
  several times, and accreted gas is more likely to remain in the
  galaxy or form stars than in the MrAGN run, in which a significant
  fraction of accreted gas ends up being ejected.}
        {\label{nacc_combined}}
\end{figure}

The fraction of the $z=0$ total galactic baryonic mass
($M_{gas}$+$M_{*}$ within $r_{\rm{g}}$) of m0163 that is made up of
accreted stars (formed in separate halos which have merged with the
main progenitor, as opposed to being formed in situ) is $\sim$ 63\% in
the MrAGN run and $\sim$ 33\% in the NoAGN run. These fractions are
shown in the far left bars of the top row of Figure
\ref{nacc_combined}. Accreted stars make up a larger fraction of the
final baryonic mass of the MrAGN galaxy than of the NoAGN galaxy
because of the reduction of in situ star formation in the former. Also
plotted is the total gas mass that has accreted since z $\sim$ 4
divided by the total baryonic mass at $z=0$, in bins of the number of
times the gas particle accreted onto the galaxy. This is further
subdivided into the fates of the accreted gas particles. Gas in the
NoAGN run is more likely to be accreted several times and is also more
likely to remain in the galaxy in the form of gas or stars. In the
MrAGN galaxy, gas tends to accrete fewer times, and the gas that is
accreted is more likely to be outside of the galaxy by $z=0$. The
fraction of total accreted gas that is ejected from the galaxy by
$z=0$ in the MrAGN run is 67.1\% as compared with 31.4\% in the NoAGN
run.

For m0329, the fraction of the $z=0$ baryonic mass that was accreted
as stars is again larger for the MrAGN galaxy than the NoAGN galaxy
($\sim$ 53\% versus $\sim$ 24\%). Also apparent is that practically
none of the accreted gas remains as gas within the MrAGN galaxy at
$z=0$, whereas some of this gas remains in the NoAGN run, while a
larger fraction is turned into stars.  In the MrAGN run, 32.6\% of all
accreted gas is ejected by $z=0$, while only 10.7\% is ejected in the
NoAGN run. These trends continue for m0501, which again has a larger
accreted stellar fraction in the MrAGN run ($\sim$ 39\% versus $\sim$
24\%). There is a slight preference for more recycling events in the
NoAGN galaxy, and again none of the accreted gas in the MrAGN galaxy
remains within the galaxy by $z=0$. 20.8\% of accreted gas is ejected
in the MrAGN run, while the rest is converted into stars. This figure
is relatively small when compared with the 14.4\% in the NoAGN run,
but this is because m0501 never has an opportunity to accrete any more
gas that would be affected by future AGN feedback.

\subsubsection{Recycling and Ejection Timescales}

\begin{figure}
\centering
  \includegraphics[width=\columnwidth]{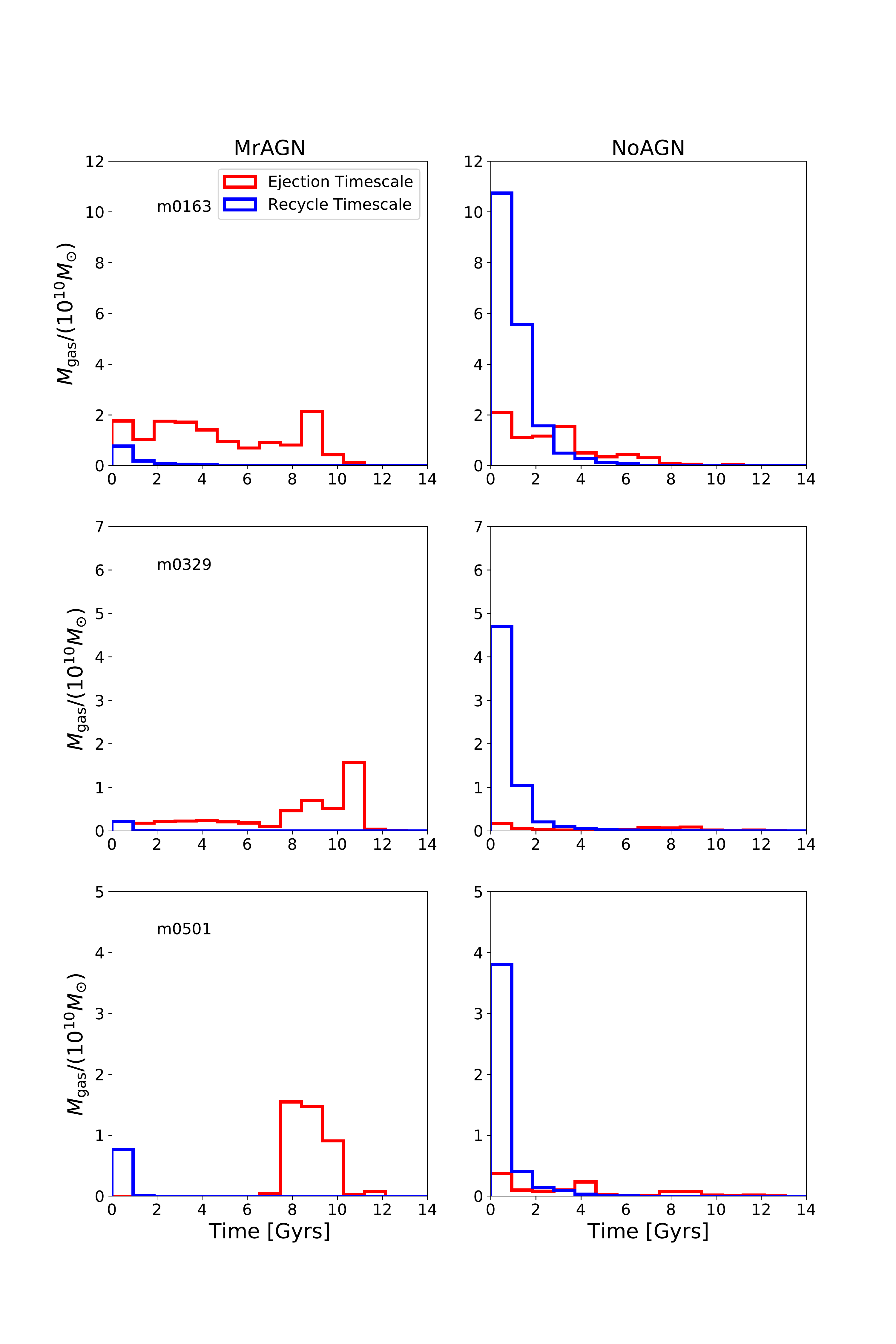}
  
  \caption[Timescales for recycling and ejection events in case studies]
  {Distribution of timescales of recycling and ejection events
    for our case study galaxies. Top row: m0163. Middle row:
    m0329. Bottom row: m0501. Left panels: MrAGN runs. Right panels:
    NoAGN runs. The blue histograms show the distribution of the 
    recycling timescale, the time spent outside the galaxy by gas that has 
    been ejected past the galaxy radius until it later came back. The red 
    histograms show the distributions of the time since ejection for gas particles that
    have exited the galaxy at the galaxy radius and have not yet returned.
    There are more recycling events in the NoAGN runs. Recycled gas
    particles in higher mass NoAGN galaxies like m0163 tend to have
    longer recycling timescales than those in lower mass
    galaxies. Ejected gas particles in lower mass MrAGN galaxies like
    m0501 are more likely to have been ejected at early times by a
    powerful bout of AGN feedback than those in higher mass galaxies.}
          {\label{timescalehist_combined}}
\end{figure}

Figure \ref{timescalehist_combined} shows distributions of timescales
of gas recycling and gas ejection events for our case study
galaxies. We define the recycling timescale as the amount of time 
between ejection of a gas particle in a wind and re-accretion into a galaxy.
Every time a gas particle crosses a shell at $\rgal$,
the time is recorded. The time is recorded again if it flows back into
the galaxy. This may happen multiple times for a single gas particle,
each of which is recorded as a separate recycling event and is
included in the distribution. If the gas particle outflows but never
comes back, the ejection timescale is recorded as the age of the
universe minus the time of ejection. The top row shows us that in the
MrAGN run of m0163, there is far less recycling on short timescales
than in the NoAGN counterpart. This appears to be because of a few
very long-term (seemingly permanent) ejection events that occurred
8-10 Gyrs ago. Those gas particles were then not available to be
recycled by stellar and supernova feedback processes as in the NoAGN
run. At later times (ejection timescales $\lesssim$2 Gyrs), the
ejection distributions look very similar. This is most likely due to
random motions of gas particles in both runs which have not had time
to return to the galaxy.

The NoAGN run of m0329 also has many more recycling events than its
MrAGN counterpart. This time, however, there are consistently more
ejection events of longer timescales in the MrAGN case, and these are
not only gas particles outflowing at early times as for m0163. The AGN
activity more consistently removes material throughout this galaxy's
history, again removing gas that would have contributed to the
short-timescale recycling events due to stellar and supernova
feedback. We also see the preference for shorter timescale recycling
in the NoAGN galaxy as compared with m0163, which is responsible for
the relatively constant inflowing recycling fraction seen in the
middle row of Figure \ref{domination_combined}.

A cumulative gas mass of $\sim4\times10^{10}M_{\odot}$ in the MrAGN
run of m0501 is ejected in a series of outflows between 7 and 12 Gyrs
ago. The NoAGN run is dominated by recycling events, with far fewer
gas particles being ejected (<$10^{10}M_{\odot}$) relative to being
recycled on short timescales (>$4\times10^{10}M_{\odot}$).

\subsubsection{Recycling and Ejection Displacements}
\begin{figure*}
\centering
  \includegraphics[width=0.65\textwidth]{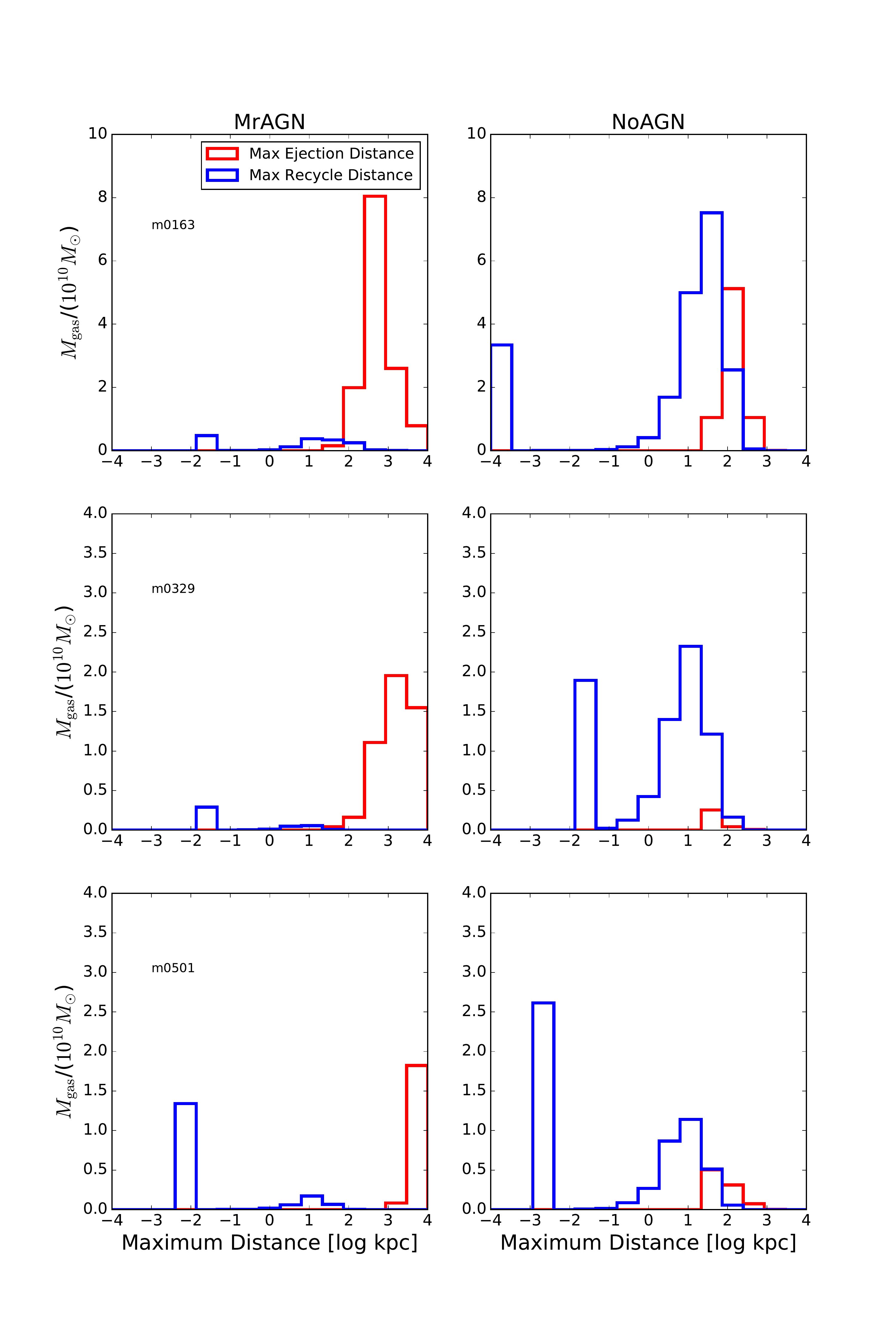}
  
  \caption[Displacements for recycling and ejection events in case studies]
  {Distribution of displacements for recycling and ejection
    events in our case study galaxies. Displacement is measured from the 
    particle's initial radius outside of the galaxy radius, recorded when the 
    particle is first tagged as outflowing. Top row: m0163. Middle row:
    m0329. Bottom row: m0501. Left panels: MrAGN runs. Right panels:
    NoAGN runs. The blue histograms show the distribution of the
    maximum distance traveled by gas particles that have been ejected
    past the galaxy radius and later come back. The red histograms show the
    distributions of the maximum distance for gas particles that have
    exited the galaxy past the galaxy radius and have not yet
    returned. Recycled particles which only ever travel back inwards
    from their initially recorded radius outside the galaxy are
    counted in our lowest bin. Gas particles ejected by AGN feedback
    travel much farther than those ejected by stellar and supernova
    feedback, and this trend becomes more extreme towards lower halo
    masses.}  {\label{rmaxhist_combined}}
\end{figure*}

Figure \ref{rmaxhist_combined} shows the distribution of maximum 
displacements of gas particles, again both for completed recycling events 
and ejection events in which the particle is still outside of the galaxy at the 
end of the simulation. When a particle is first tagged as outflowing, 
its current radius is recorded as the initial radius (usually just outside of 
$r_{\rm{g}}$). We keep track of its subsequent radius and store the 
maximum radius reached during a given recycling or ejection event. We 
then subtract the initial radius to get the maximum displacement 
experienced by a given gas particle. For m0163, we see once again 
that there is far more mass ejected ($\sim10^{11}M_{\odot}$) than 
recycled (<$10^{10}M_{\odot}$) in the MrAGN run, due to a larger 
number of gas particles which are part of outflows at early times and 
never re-accrete onto the galaxy. It is also evident that the outflowing 
gas particles removed by AGN feedback travel much larger distances 
than those removed by stellar and supernova feedback. In some 
cases, gas particles travel several Mpc, while in the NoAGN run only 
the very tail of the distribution reaches $\sim$ 1 Mpc.

The distribution of maximum displacements for gas particles in the MrAGN 
run of m0329 is even more extreme than for m0163, with gas particles 
being driven out beyond the zoom region of our simulation. While we keep 
track of these distances, here we cap the particles' displacements at the 
boundaries of our zoom region, which ranges from 5 to 10 Mpc for 
our 24 halos. Outside of this zoom-in region only the low resolution dark 
matter particles are populated to approximate the long-distance tidal force 
\citep{Oser2010}, and therefore the wind particles pushed outside of the 
zoom-in region can travel even further. However, outflows can be stalled 
due to intergalactic medium pressure and then fall back to the galaxy halo 
in reality. This is the caveat that arises from the limited size of the zoom-in 
region, however, its effect on our infall rate measurement is believed to be 
small, given that the displacement of the recorded recycling events is two 
orders of magnitude smaller than the scale of the zoom-in region. Ejected 
particles in the NoAGN case can travel several hundred kpc, but only the 
extreme tail of the distribution exhibits radii larger than this. In both cases, 
the recycled material is restricted to these smaller displacements as well.

In m0501, with the halo depleted and the surrounding filament destroyed 
there is nothing to stop ejected gas from traveling tens of Mpc in the 
MrAGN run. In contrasst, ejected particles in the NoAGN run mostly only 
travel 100-200 kpc. In both runs, recycled gas particles are restricted to 
smaller radii, as is expected.

\subsection{Broad Trends in Galaxy and Outflow Properties}

In the last section, we focused on three individual example galaxies and 
found that the strength of gas outflows and inflows depend on halo mass. 
We now turn to the broad trends and scaling relations for galaxy and 
outflow properties that can be extracted from our complete suites of 
galaxies. In the following plots, each point represents a single galaxy at 
the specified redshift. Some galaxies do not have a matched pair or reliable 
center by $z=3$, or in some cases $z=2$, and so these are not plotted at 
high redshift.

\subsubsection{Galaxy Properties}

\begin{figure*}
  \includegraphics[width=1.0\textwidth]{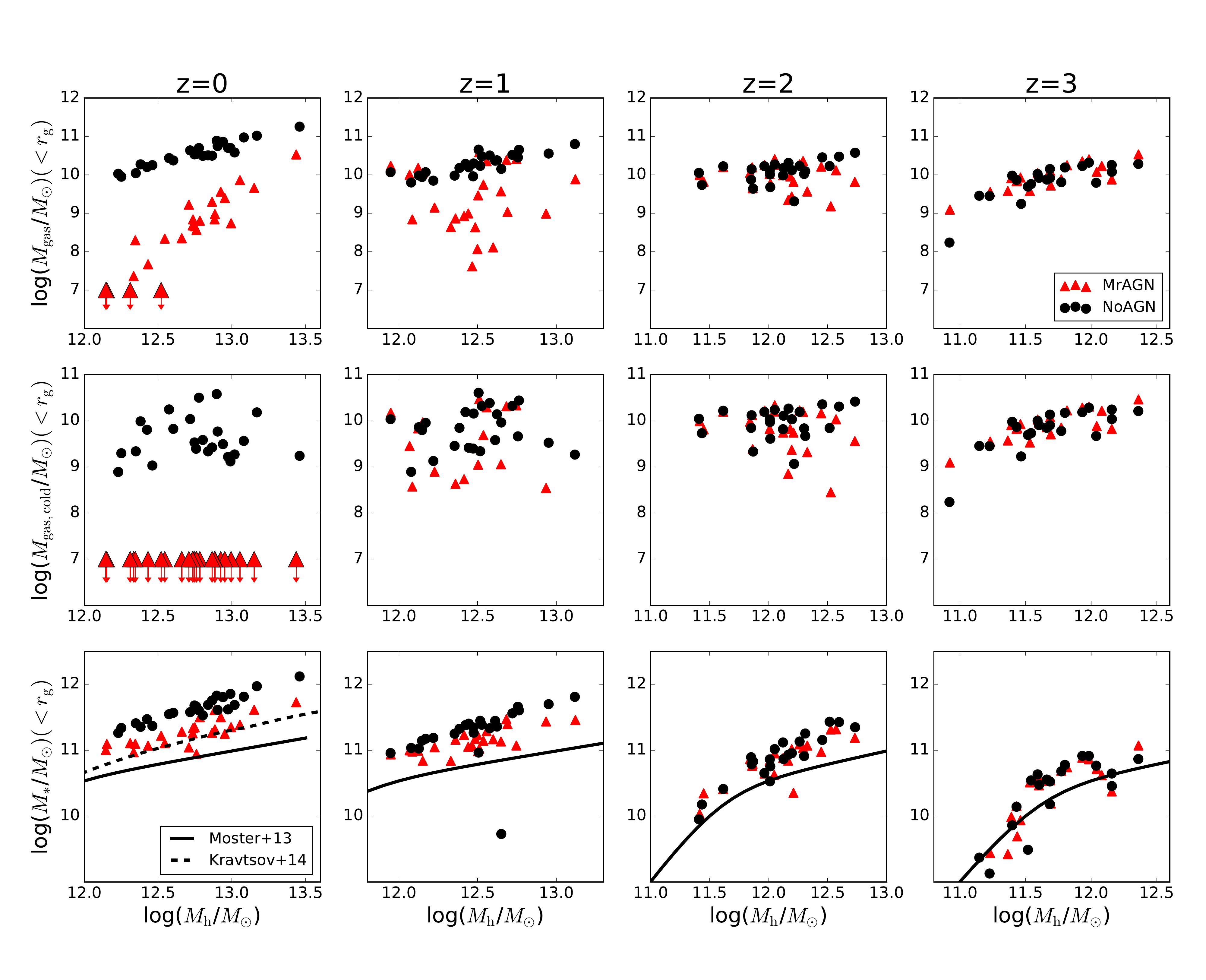}
  
  \caption[Gas and stellar content versus halo mass]{Gas and stellar 
  content of all galaxies at $z=0$ vs. halo mass. Top row: Total mass of 
  gas within 10\% of the virial radius, $\rgal$. Middle Row: Cold gas mass 
  (T<2$\times10^{4}$ K) within $\rgal$. Bottom Row: 
  Stellar mass. The solid black curves are the abundance matching 
  estimates from \citet{Moster2013} and the dashed black curve is the $z=0$
    abundance matching estimate of \citet{Kravtsov2014}. Black circles
    denote galaxies in the NoAGN run, while red triangles denote
    galaxies in the MrAGN run. Quantities plotted at $10^{7}M_{\odot}$
    with downward arrows are galaxies for which the y-axis value is
    0. By $z=0$, all MrAGN galaxies have significantly less gas (and
    in some cases no gas), and all have been cleared of cold gas
    completely. All MrAGN galaxies also have lower stellar masses, as
    was seen in the individual galaxy histories shown above.}
          {\label{various_mhalo}}
\end{figure*}

Figure \ref{various_mhalo} shows the gas and stellar mass of all of
our galaxies versus halo mass for four different redshifts. The full
cold gas mass and stellar mass histories of our three case study
galaxies can be found in Figures \ref{flowplot_163},
\ref{flowplot_329} and \ref{flowplot_501}. At $z=3$, before most major
AGN activity (when supermassive black holes are still fairly small),
galaxies in both runs have very similar gas content and fall on a
fairly tight relation between gas mass and halo mass. At $z=2$, AGN
feedback begins to kick in and MrAGN galaxies start to drop below the
NoAGN relation. By $z=1$, much of the gas affected by star formation
and stellar feedback has fallen back into the galaxies, while in the
MrAGN runs, much of this gas is lost. At $z=0$, galaxies in the two
runs fall on distinct relations, with some of the lowest halo mass
galaxies having no gas left within $\rgal$.  The middle row tells
a similar story about the cold gas content (defined as gas with
T<2$\times10^{4}$K), except for two important differences. The
relationship between cold gas and halo mass for NoAGN galaxies has
more scatter than the total gas mass to halo mass relation. Secondly,
by $z=0$ none of the MrAGN galaxies has any cold gas left at all.
Finally, the bottom panels show the stellar mass of the central galaxy
versus the halo mass, and reveal a more gradual separation of the two
relations due to the permanent removal and/or heating of cold gas
causing the buildup of stars in the MrAGN runs to lag behind that of
the NoAGN runs. The solid black curves are an estimate of the
stellar-mass halo mass relation from the abundance matching analysis
of \citet{Moster2013}. The dashed black curve is the $z=0$ abundance
matching estimate of \citet{Kravtsov2014}. Our MrAGN galaxies are a
very good match to the \citet{Kravtsov2014} estimate, which used
improved photometric techniques and took into account intracluster light 
to measure stellar masses.

\subsubsection{Inflow and Outflow Properties}

\begin{figure*}
  \includegraphics[width=1.0\textwidth]{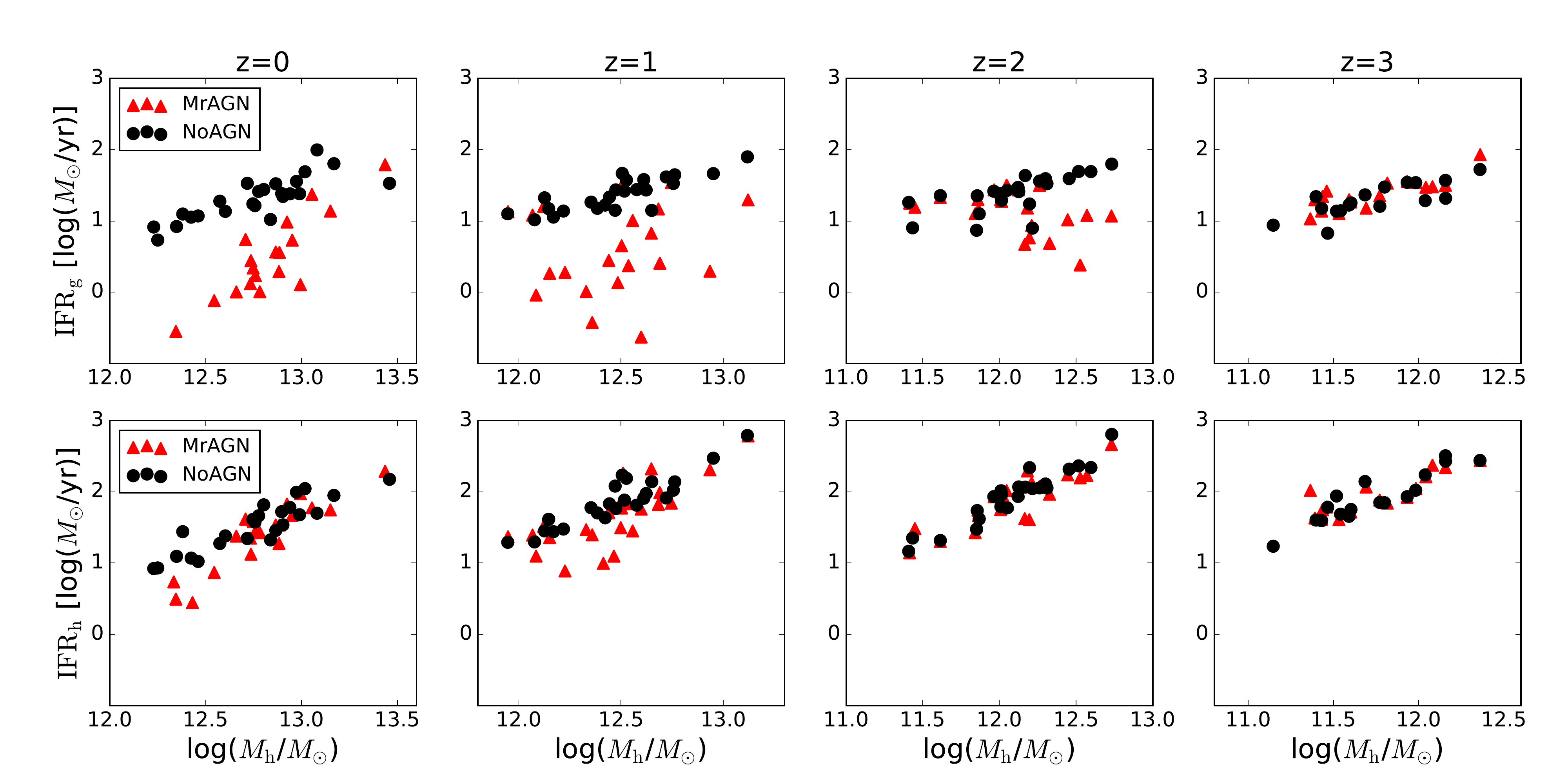}
  
  \caption[Inflow rate versus halo mass]{The inflow rate of gas at the 
  specified redshift. Top row:  Inflow across the galaxy radius. Bottom 
  row: Inflow across the halo radius. Black circles denote galaxies in 
  the NoAGN runs, while red triangles denote galaxies in the MrAGN 
  runs. The inflow rate at the galaxy radius for MrAGN galaxies is 
  decreased by as much as 1.5 dex by $z=0$. We see evidence of 
  downsizing in the initial decrease of inflow for high mass galaxies at 
  z=2 when their AGN begin to switch on. At the halo radius, the inflow 
  rate is only suppressed for lower mass halos.}  {\label{inflow_rate}}
\end{figure*}

Next we turn to inflow and outflow properties. Again, the full inflow
and outflow histories of our case study galaxies have been presented
above; here we present snapshots of these quantities at the
specified redshifts for our entire sample of galaxies. Figure
\ref{inflow_rate} illustrates that at $z=3$, the inflow rates for MrAGN
and NoAGN galaxies are very similar and dependent on halo mass. By
$z=2$, the higher mass MrAGN galaxies begin to have inflow suppressed by
$\gtrsim$1 dex as their AGN turn on. By $z=1$, this phenomenon is
widespread and the inflow rate is again broadly dependent on halo
mass. At $z=0$, the highest mass galaxies are again accreting at a rate
comparable to their NoAGN counterparts, while the rest of the MrAGN
galaxies continue to have their inflow rates suppressed by up to 1.5
dex. This again shows the importance of preventative feedback,
especially in lower mass galaxies.

At $\rhalo$ the evolution is very similar, although the
suppression of inflow at this larger scale is much less pronounced. At
$z=1$ and $z=0$ it is the lower mass MrAGN galaxies that have their inflow
suppressed (although now by $\sim$ 0.5 dex), while more massive
galaxies have inflow rates closer to their  NoAGN counterparts. However, as 
shown in Table 1, the cumulative inflowing mass at $\rhalo$ expressed as a 
fraction of the universal baryon fraction times the halo mass at $z=0$ is 
smaller in MrAGN galaxies than in NoAGN galaxies, even at larger 
masses. Still, the effect is more pronounced at lower masses. At both
scales there is not much evolution in the inflow rates of NoAGN
galaxies. At both radii we see evidence of downsizing: AGN
feedback occurs in more massive galaxies first, as evidenced by the
initial decrease in inflow rate for the most massive galaxies at
z $\sim$ 2, and more strongly affects lower mass galaxies at later
times.

\begin{figure*}
  \includegraphics[width=1.0\textwidth]{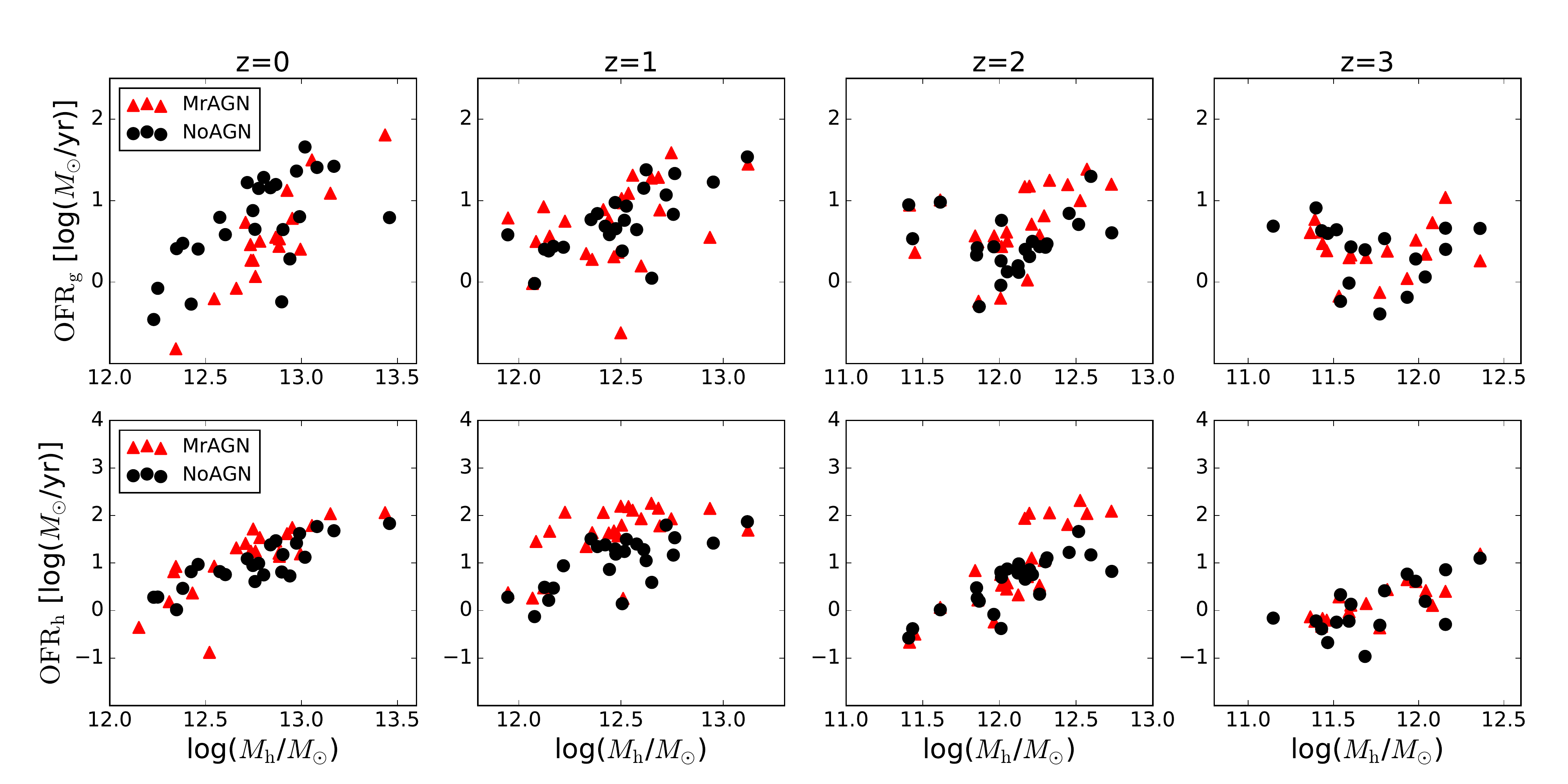}
  
  \caption[Outflow rate versus halo mass]{The outflow rate of gas at the 
  specified redshift. Top row: outflow across the galaxy radius. Bottom 
  row: outflow across the halo radius. Black circles denote galaxies in the 
  NoAGN runs, while red triangles denote galaxies in the MrAGN runs. The 
  outflow rate at the galaxy radius for MrAGN galaxies is not appreciably
  different from that of NoAGN galaxies. At the halo radius, we see an
  elevated outflow rate for MrAGN galaxies, mainly at $z=2$ and
  $z=1$. We again see evidence for downsizing, with the most massive
  galaxies experiencing enhanced outflows at the halo radius before lower
  mass galaxies.}  {\label{outflow_rate}}
\end{figure*}

Figure \ref{outflow_rate} is the same as Figure \ref{inflow_rate}, but
now for the outflow rate. At $\rgal$ the outflow rates of MrAGN
galaxies at all redshifts are comparable to the outflow rates of NoAGN
galaxies. At $\rhalo$, however, we see elevated outflow rates,
first for more massive galaxies at $z=2$, then for most galaxies by
$z=1$. The outflow rates of MrAGN galaxies remain slightly elevated by
$z=0$ for many galaxies. This confirms what was shown earlier for the
case studies; outflowing gas driven by AGN feedback is less likely to
fall back into the galaxy and is more likely to travel out past the
virial radius.

\begin{figure*}
  \includegraphics[width=1.0\textwidth]{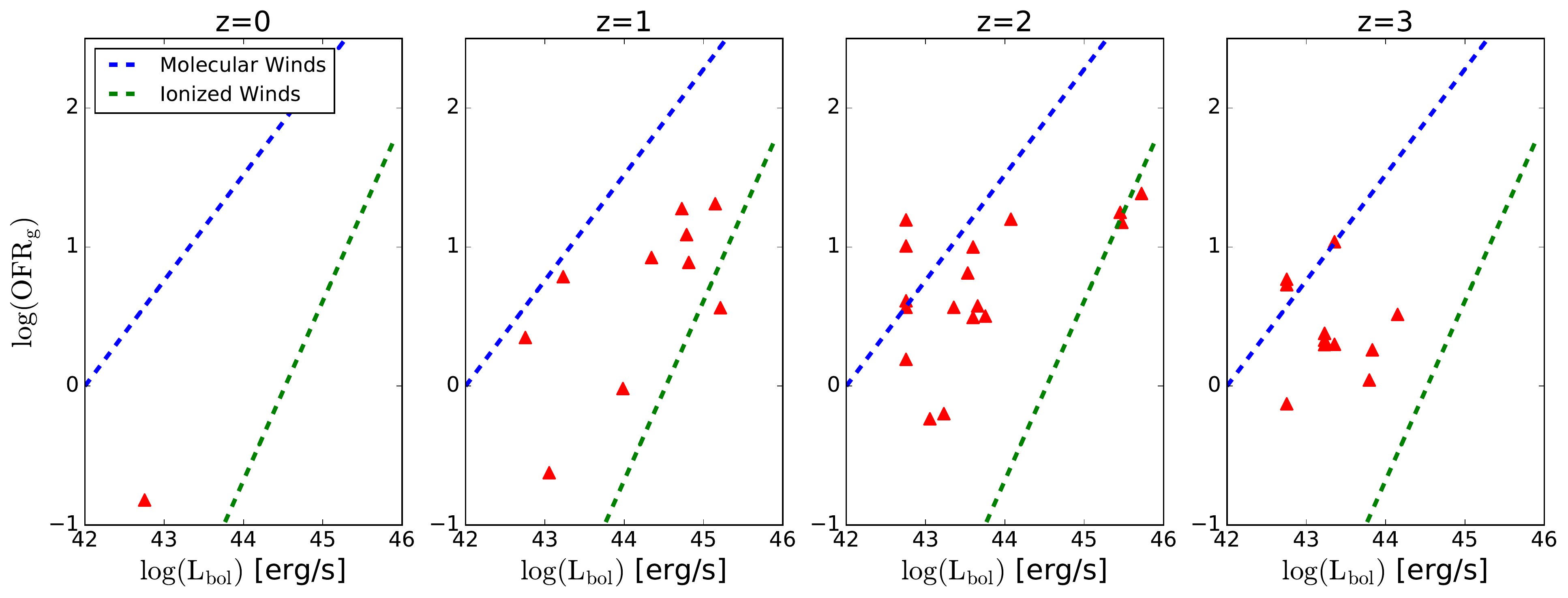}
  
  \caption[Outflow rate versus bolometric AGN luminosity]{The outflow
    rate of gas across galaxy radius $\rgal$ versus the bolometric luminosity of the
    galaxy's AGN. MrAGN galaxies are shown as red triangles. The
    dashed blue and green lines are scaling relations found for
    AGN-driven molecular and ionized outflows, respectively, and
    presented in \citet{Fiore2017}. These scaling relations were
    derived for outflows observed at a number of different radii, and
    there is large uncertainty on some of the values used to derive
    the relations, but our results appear to be in qualitative
    agreement with the observed scaling relations found in the
    literature and compiled in \citet{Fiore2017}.}
          {\label{outflow_rate_lbol}}
\end{figure*}

Figure \ref{outflow_rate_lbol} again shows the outflow rates of our
MrAGN galaxies at four different redshifts, this time versus the
bolometric luminosity of their AGN. Higher luminosity AGN tend to be
correlated with larger outflow rates. The blue and green dashed lines
represent scaling relations presented in \citet{Fiore2017} for
molecular and ionized winds, respectively, for a compilation of
observed AGN galaxies with detectable outflows. While the measurements
of outflow velocity probe different radii in the galaxies used to
define these relationships, we are encouraged that our MrAGN galaxies
occupy a realistic portion of the OFR-$L_{\rm{bol}}$ plane, and thus
qualitatively agree with observed winds. \citet{Yesuf2017} have also 
observed winds in elliptical galaxies with AGN at $z\sim0.1$ and found 
velocities as large as $\sim680$ km/s, also in qualitative agreement with 
our results.

\begin{figure*}
  \includegraphics[width=1.0\textwidth]{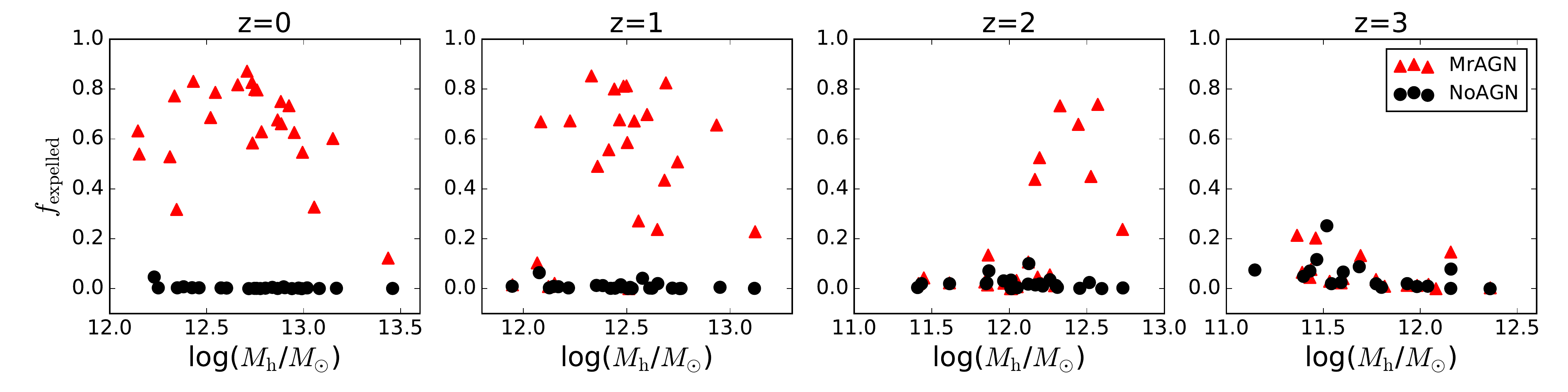}
  
  \caption[Expelled fraction versus halo mass]{The fraction of the cumulative mass which has crossed a
    shell at the galaxy radius and which has subsequently crossed a shell
    at the halo radius by the specified redshift. Black circles denote
    galaxies in the NoAGN runs, while red triangles denote galaxies in
    the MrAGN run. We see that a much larger fraction of gas that
    outflows past the galaxy radius eventually crosses the halo radius in the
    MrAGN runs.}  {\label{fexpel}}
\end{figure*}

Figure \ref{fexpel} further quantifies the fraction of gas that has
crossed a shell at $\rgal$ that eventually crosses a shell at
$\rhalo$ as a function of halo mass. Whenever a gas particle is
tagged as outflowing past $\rgal$, its radius is tracked and, if
it later crosses $\rhalo$, it is considered expelled. At each
redshift shown, we have divided the cumulative mass that has been
expelled by the cumulative mass of gas that has been considered
outflowing past $\rgal$ up until that redshift. These cumulative masses may include
certain gas particles multiple times if they are recycled. At $z=3$,
when galaxies are small, this ratio is similar for MrAGN galaxies and
NoAGN galaxies. At $z=2$, we see the effect of AGN turning on in higher
mass galaxies as in Figures \ref{inflow_rate} and
\ref{outflow_rate}. At later redshifts, once AGN feedback kicks in for
all galaxies, this ratio is much larger for MrAGN galaxies than NoAGN
galaxies, as gas particles ejected by AGN feedback tend to travel much
farther. Interestingly, there does not appear to be much of a
consistent trend with halo mass for this quantity.

\begin{figure*}
  \includegraphics[width=1.0\textwidth]{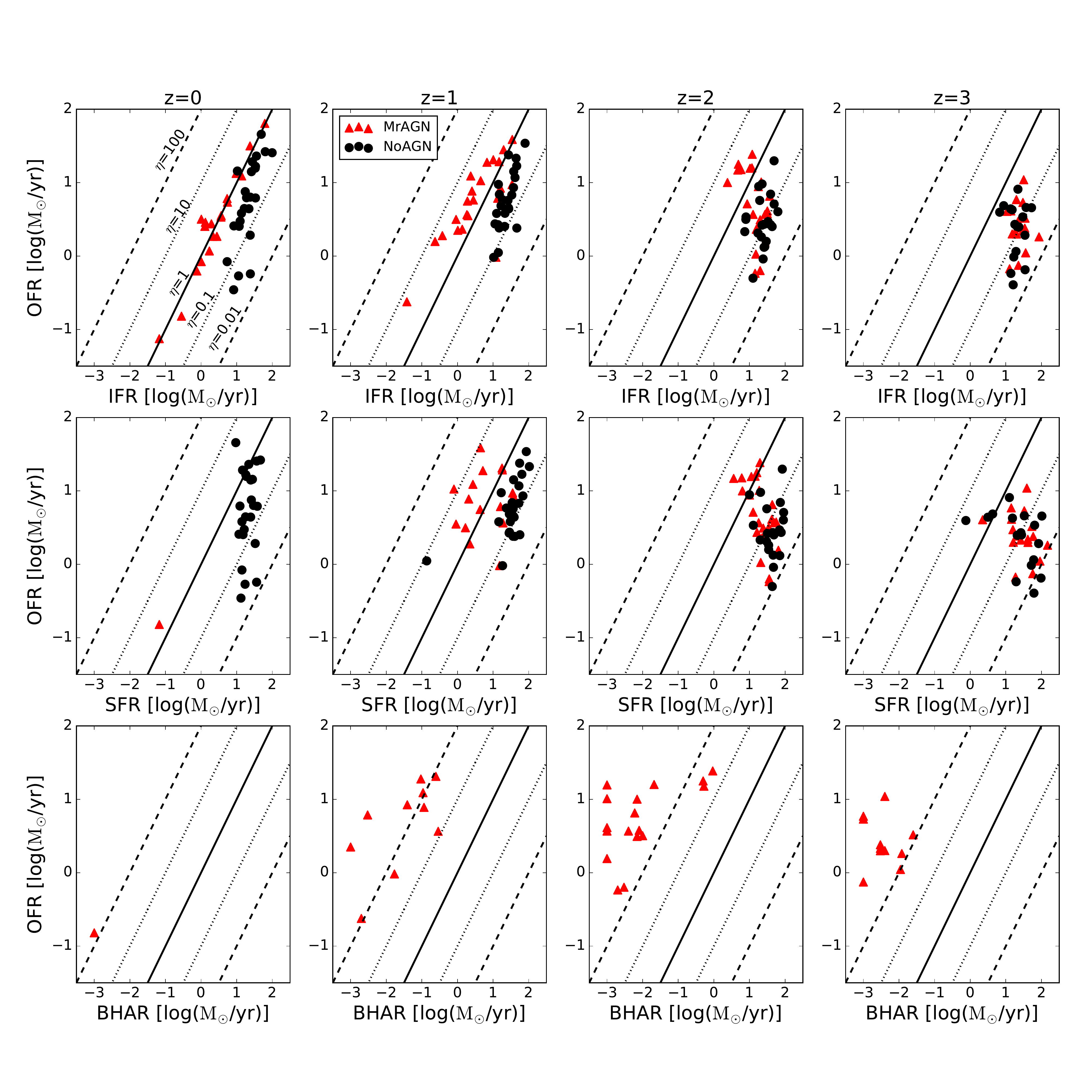}
  
  \caption[Mass loading factors]{Outflow rate at the galaxy radius versus: inflow rate at
    the galaxy radius (top row), star formation rate (middle row), and black hole accretion rate (bottom row,
    for MrAGN runs). Black circles denote galaxies in the NoAGN runs,
    while red triangles denote galaxies in the MrAGN run. Black lines
    specify constant mass loading factors, $\eta$. Top row:
    $\eta$=outflow rate/inflow rate. Middle row: $\eta$=outflow rate/star formation rate. Bottom row:
    $\eta$=outflow rate/black hole accretion rate. At $z=2$ and $z=1$ the MrAGN galaxies tend to
    have higher loading factors than the NoAGN galaxies, although by
    $z=0$, there is only one MrAGN galaxy that still exhibits star
    formation or black hole accretion.}  {\label{loading_various}}
\end{figure*}

In Figure \ref{loading_various}, we examine different types of loading
factors by plotting outflow rates versus inflow rates, both measured
at $\rgal$ (top row), SFR (middle row) and BHAR (bottom row).  In
the top row, we consider outflow rate versus inflow rate. At $z=3$,
the two runs sit on top of each other at loading factors less than
unity. By $z=2$, the loading factors of MrAGN galaxies tend to be
larger by as much as ten times, and by $z=1$ as much as a hundred
times, with some galaxies having outflow rates larger than their
inflow rates. This is due to a combination of both higher outflow
rates and lower inflow rates for MrAGN galaxies. By $z=1$, almost
every MrAGN galaxy has a loading factor greater than unity, while the
NoAGN galaxies all sit below this line. Now, however, this difference
is almost entirely due to the suppressed inflow rates of MrAGN
galaxies. Finally, at $z=0$, it appears that outflow and inflow are
regulating each other in the case of the MrAGN galaxies, while the
NoAGN galaxies hit a floor in inflow rate, as stellar and supernova
feedback and gravitational heating do not suppress inflow as
efficiently as AGN feedback does.  Turning to outflow rate versus star
formation rate, at $z=3$, 2 and 1, this loading factor follows a
similar pattern as outflow rate/inflow rate. At $z=2$, the outflow
rates of MrAGN galaxies are higher than NoAGN galaxies, while their
SFRs are lower. At $z=1$, the difference in loading factor is due
almost entirely to the decreased SFRs of MrAGN galaxies. At $z=0$, the
one MrAGN galaxy still forming stars has a much lower SFR than any of
the NoAGN galaxies, but also a very low outflow rate.  Finally, the
bottom panels show that the outflow rates of MrAGN galaxies are 100
times or more greater than their black hole accretion rates at $z=3$,
although this value falls slightly with redshift. There are also fewer
galaxies with appreciable black hole accretion rates as we go from
$z=2$ to $z=0$.

\begin{figure*}
  \includegraphics[width=1.0\textwidth]{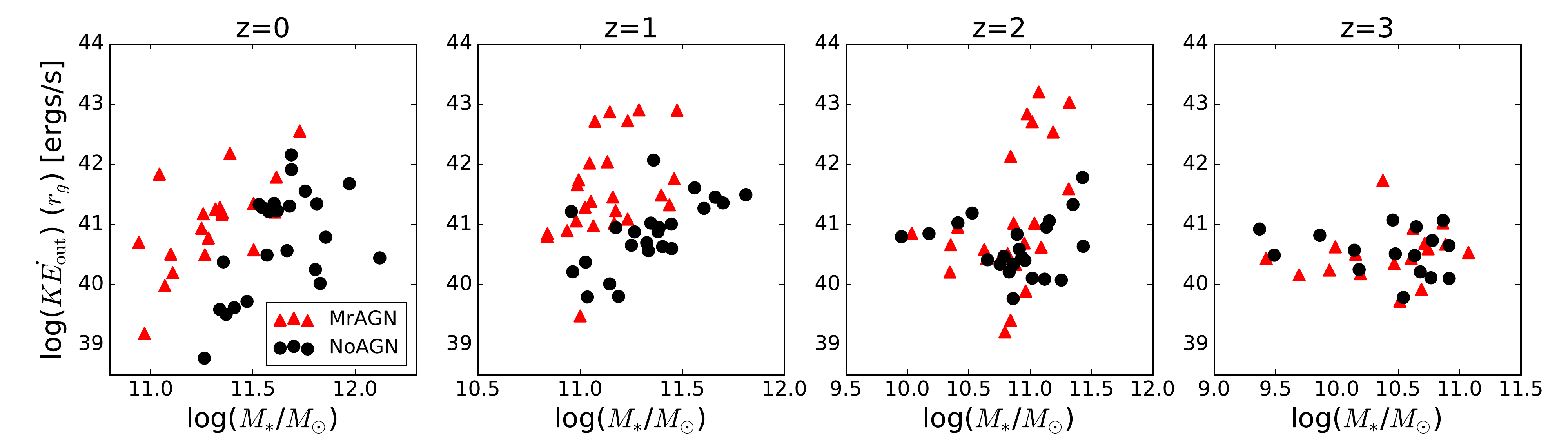}
  
  \caption[Kinetic energy versus stellar mass]{Kinetic energy outflow rate of all gas particles which have crossed a shell at the galaxy radius since the last timestep vs. stellar mass. The kinetic energy outflow rate is calculated by summing the total kinetic energy of all of the outflowing particles and dividing by the time between snapshots. Black circles denote galaxies in the NoAGN runs, while red triangles denote galaxies in the MrAGN run. At $z=3$ and $z=0$, the total kinetic energy of outflowing particles is very similar between the two runs, although the stellar masses of MrAGN galaxies are characteristically lower. At $z=2$ and $z=1$, MrAGN galaxies have more kinetic energy in outflows by as much as two orders of magnitude.}
          {\label{ke_stellar}}
\end{figure*}

\begin{figure*}
  \includegraphics[width=1.0\textwidth]{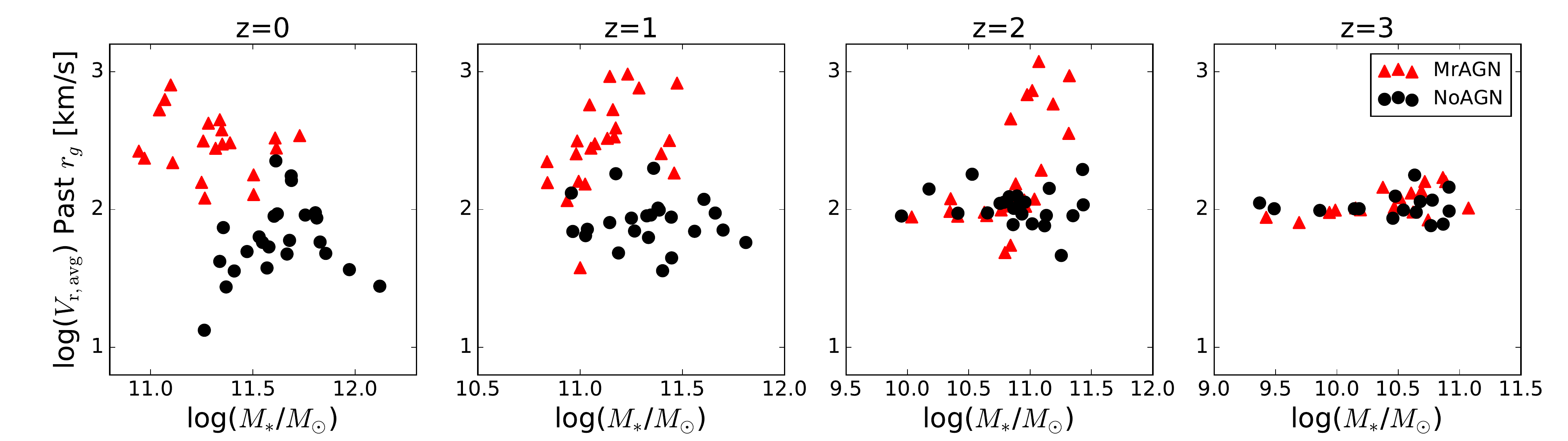}
  
  \caption[Radial velocity versus stellar mass]{Average radial velocity of all gas particles just crossing a shell at the galaxy radius at the specified redshift vs. stellar mass. Black circles denote galaxies in the NoAGN runs, while red triangles denote galaxies in the MrAGN run. For redshifts of 2 and less, the average radial velocity of gas outflowing past the galaxy radius in MrAGN galaxies is much higher than that of gas in NoAGN galaxies.}
          {\label{vr_stellar}}
\end{figure*}

In Figures \ref{ke_stellar} and \ref{vr_stellar}, we examine, very
broadly, the kinematics of outflowing material. Figure
\ref{ke_stellar} shows the kinetic energy outflow rate for particles
which have crossed a shell at $\rgal$ since the last timestep
versus stellar mass. The kinetic energy outflow rate is calculated by
summing the total kinetic energy of all of the outflowing particles
and dividing by the time between snapshots. At $z=3$, the MrAGN
galaxies and NoAGN galaxies are very similar, but by $z=2$, some of
the MrAGN galaxies have much more kinetic energy in outflowing
particles. By $z=1$ this trend is even clearer, and the MrAGN galaxies
have started to become displaced to lower stellar masses as well. At
$z=0$, the kinetic energy in outflowing particles is once again
similar in the two runs, but the outflows have left their mark on the
MrAGN galaxies, which have ssmaller stellar masses.

Figure \ref{vr_stellar} instead illustrates the average radial
velocity of the same outflowing gas considered in Figure
\ref{ke_stellar}. At $z=3$, 2 and 1, the trends are quite similar to
those seen in the last figure. Large outflow velocities driven by AGN winds begin to be
seen at $z=2$ and at $z=1$, and almost all MrAGN galaxies have outflowing
material with significant radial velocities. At $z=0$, whereas the total
kinetic energy of outflowing particles for MrAGN and NoAGN galaxies
were very similar, the radial velocities of those outflowing gas
particles are characteristically larger in the MrAGN run. The
similarity in kinetic energies is due to the fact that there is less
mass in outflowing gas in the MrAGN galaxies; much more gas has
already been removed permanently by AGN feedback. In the NoAGN run,
the same gas is allowed to outflow and re-accrete. Higher radial
velocities of MrAGN galaxies at $z=0$ are mostly due to the higher gas 
temperature where outflows are roughly balanced by gas inflows with 
loading factor $\eta \sim 1$ (see Figure~\ref{loading_various}), except 
a few galaxies with $\eta > 1$, which had AGN outburst in the recent
past.
%We will now
%investigate gas recycling in more detail.

\subsubsection{Gas Recycling}

\begin{figure*}
  \includegraphics[width=1.0\textwidth]{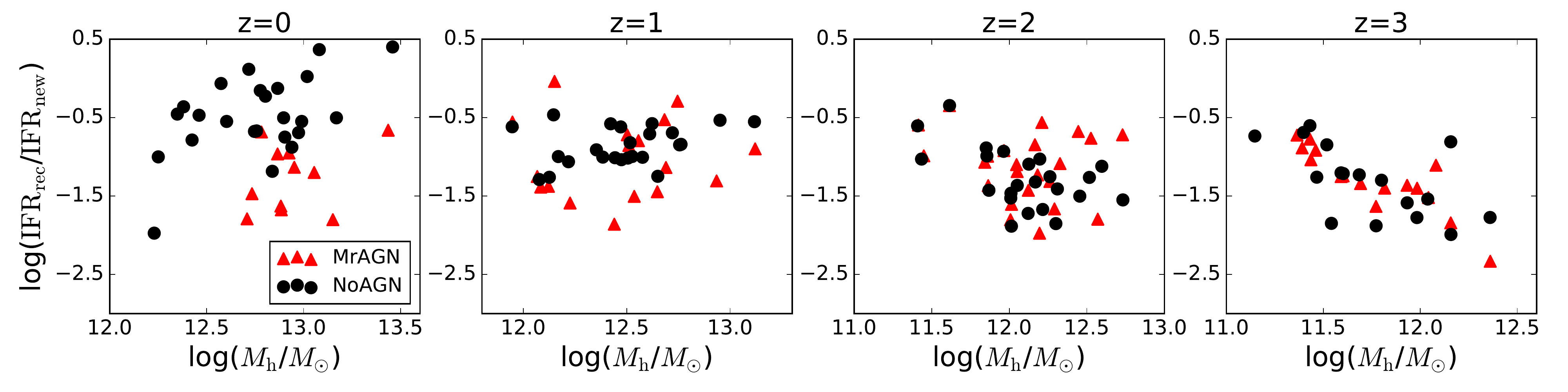}
  
  \caption[Contribution of recycled material to inflow]{Instantaneous inflow rate at the galaxy radius composed of recycled accretion divided by the inflow rate of new gas at the specified redshift. Black circles denote galaxies in the NoAGN runs, while red triangles denote galaxies in the MrAGN run. While MrAGN galaxies are almost always dominated by new accretion, NoAGN galaxies at $z=0$ experience more recycled accretion, and are even dominated by recycled accretion in the large halo mass regime.}
          {\label{frec}}
\end{figure*}

Figure \ref{frec} depicts the ratio of the inflow rate of recycled
material at $\rgal$ at the specified redshift to the inflow rate
of new material. This is related to the inflowing recycling fraction
depicted for our case studies. Down to $z=1$, galaxies in both runs are
more likely to accrete a larger portion of new gas regardless of halo
mass. This remains true for galaxies in the MrAGN run to $z=0$, but
NoAGN galaxies experience more recycled accretion, with larger mass
halos being dominated by recycling.

\begin{figure*}
\centering
  \includegraphics[width=0.8\textwidth]{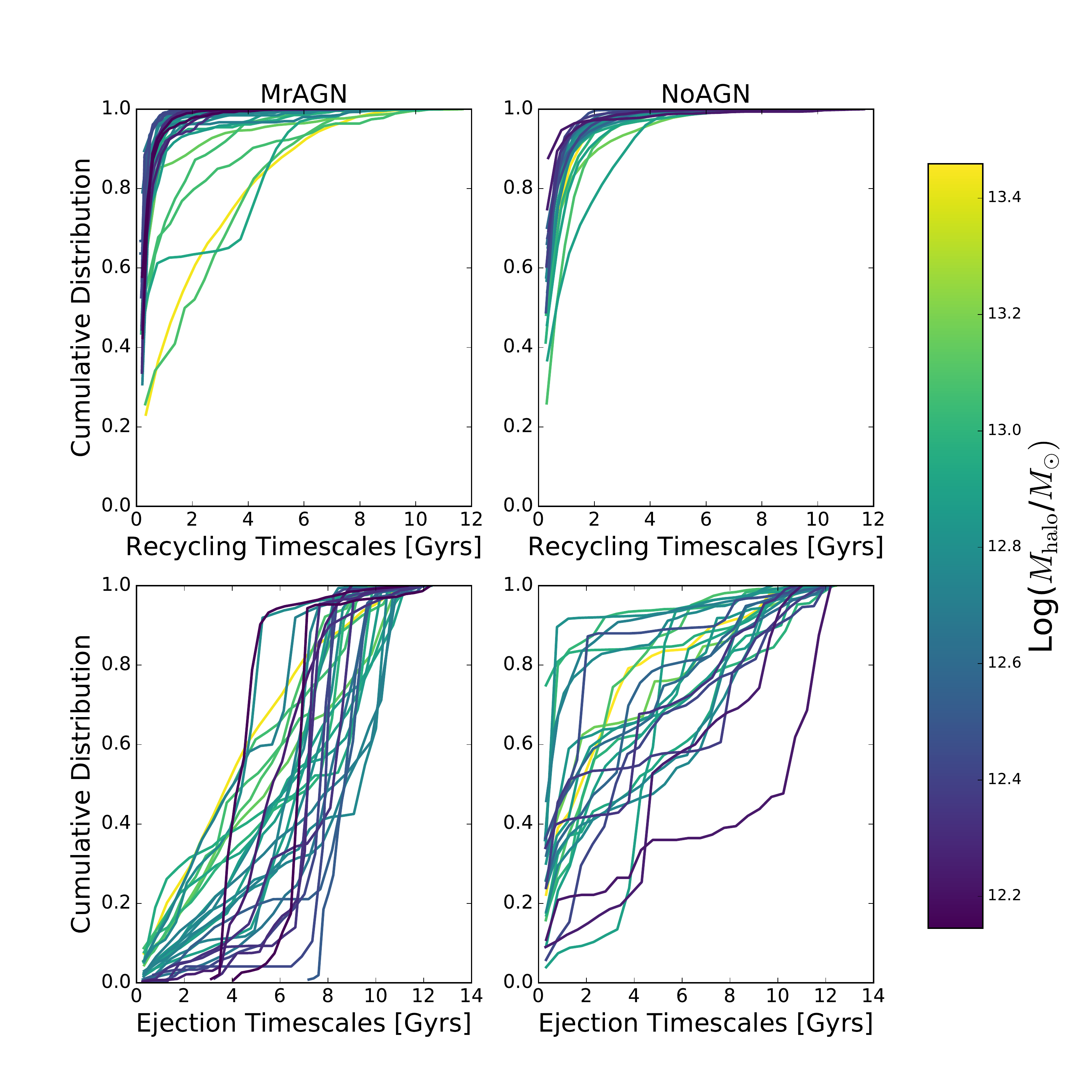}
  
  \caption[Timescale distribution of recycling events versus halo mass]{The normalized cumulative distribution of recycling (top row) and ejection (bottom row) event timescales by $z=0$, color-coded by halo mass. Purple corresponds to lower halo mass, while yellow corresponds to higher halo mass. Left panels: MrAGN galaxies. Right panels: NoAGN galaxies. MrAGN galaxies have characteristically longer recycling timescales, especially for higher mass galaxies. MrAGN galaxies also have characteristically longer ejection timescales, often the result of large, early bouts of AGN feedback. The latter trend is strongest for lower mass galaxies.}
          {\label{timescales}}
\end{figure*}

In Figure~\ref{timescales}, we have taken the histograms like those
shown in Figure~\ref{timescalehist_combined} for all of our galaxies
and turned them into cumulative timescale distributions. While we lose
information on the relative number of recycling and ejection events,
those properties are in line with those found for the example galaxies
above. The cumulative distributions are color-coded by halo mass, with
purple corresponding to the least massive and yellow to the most
massive. MrAGN galaxies have characteristically longer recycling
timescales due to the more efficient and energetic feedback. This is
especially true for more massive galaxies; while the feedback is
strong enough to lengthen recycling timescales, it is harder to remove
the gas completely from a very massive halo, resulting in gas that is
gone for a long time but nevertheless comes back. Although MrAGN 
galaxies show characteristically longer recycling timescales compared to
NoAGN galaxies overall, we note that some 
small mass MrAGN galaxies show cumulative distributions
toward shorter recycling timescales compared to NoAGN galaxies. This 
is because gas recycling happens only at early epochs with short timescales 
and the vast majority of gas is ejected for these galaxies. We also find that 
ejection timescales are characteristically longer for MrAGN galaxies; this 
time the longest timescales correspond to the less massive haloes which 
experienced strong early bouts of AGN feedback that removed gas 
permanently.

\section{Discussion}

In this section we will first compare our results with those from
other simulations, then discuss our results in the context of several
questions asked in the Introduction.

\subsection{Comparison with Other Work}

Other studies have examined the baryon cycle in simulations, although
usually these have been focused on lower mass galaxies than the ones
we study here. Despite this, we now discuss how our results compare
with some of these studies.

\citet{Oppenheimer2010} examined a set of galaxies run with
\texttt{GADGET-2} \citep{Springel2005} with various prescriptions for
stellar driven winds, tied to the galaxies' star formation rate. They
find that when galactic winds are included, the contribution of
recycled material to accretion at late times (especially z<1) is very
significant. These outflow models result in too many galaxies with
$M_{\rm *}\gtrsim2\times10^{12}M_{\odot}$ because the material that
is removed is re-accreted, and they posit that AGN feedback may
suppress this re-accretion. These results are consistent with our
large recycling fractions in the No AGN runs; even if our sample is
not cosmologically representative, we confirm their speculation that
AGN feedback can suppress late re-accretion of gas.

\citet{Hirschmann2013} examine cosmological zoom simulations using the same
stellar-driven wind model as \citet{Oppenheimer2010} and find the
same large inflow rate at late times, leading to too much in situ star
formation and galaxies which are too massive. They also found outflow
rate versus SFR mass loading factors at high redshift ($z\sim2-3$) as
large as $\sim100$, which is higher than we find in either our MrAGN
or NoAGN runs.

\citet{Uebler2014} examined a series of cosmological zoom simulations,
also run with GADGET, with masses slightly below our mass range and
overlapping with our lower mass galaxies. Much like this work, they
compared two feedback models, one with weak feedback from massive
stars, the other with strong feedback. They examined the baryon cycle
around these galaxies and found that stronger stellar feedback
produced disc galaxies with properties comparable to
observations. Their strong feedback model is similar to our NoAGN
model, and they find similar gas behavior. They find that at early
times, some gas is ejected from the galaxies permanently, due to their
shallower potential wells, while later the galaxies are dominated by
accretion of recycled gas, like ours. Their Figure 6 is the
inspiration for our Figure \ref{nacc_combined}.

\citet{Christensen2016} carried out a study of outflows in 20 field
galaxies run with the SPH code \texttt{GASOLINE}
\citep{Wadsley2004}. They looked at galaxies with halo masses of
$10^{9.5}\--10^{12}M_{\odot}$. They find that 50\% of gas that leaves
the galaxies is later re-accreted regardless of mass. Our results are
in agreement with theirs in that we find that the future recycling
fraction for our NoAGN galaxies is very similar for all galaxies, also
regardless of mass. However, we also find that the cumulative
inflowing mass of higher mass galaxies is more likely to be recycled
($\sim25-35\%$) than in lower mass galaxies (as low as $\sim7\%$). Our
results also suggest that a very common recycling timescale for gas in
our NoAGN galaxies is $\sim$ 1 Gyr, in agreement with their results.

\citet{Barai2016} implemented mechanical AGN feedback similar to ours
in a cluster-sized halo using \texttt{GADGET-3}. Although our analysis is quite
different, they come to several similar conclusions. They find bipolar
bubble-like outflows of heated gas out to hundreds of kiloparsecs,
which is very similar to the structure of our outflows. They also find
that kinetic feedback is far more efficient at affecting central gas
and quenching star formation than thermal feedback.

\citet{Angles2016} examined a suite of galaxies in the FIRE simulation
\citep{Hopkins2014}, which span a lower range of halo masses than the
ones we have looked at and also do not include AGN feedback. They find
(for galaxies with stellar feedback) that galaxies with lower halo
mass are more dominated by recycled accretion and become more so at
low redshift, while higher mass halos are dominated by new
accretion. We find that in general NoAGN galaxies are more dominated
by recycled accretion at late times, although this is much more
prevalent for the higher mass galaxies in our halo mass range.  With
our AGN feedback prescription we find that there is very little
recycled accretion onto our galaxies, even at late times, regardless
of galaxy mass.

\citet{Angles2016} also find that in the absence of AGN feedback, the
most massive galaxies have the shortest recycling timescales, although
they do find that these same galaxies have larger recycling distances
than lower mass galaxies. We find that our NoAGN galaxies are all
dominated by relatively short recycling timescales and these
timescales are longer for more massive halos. Recycled gas remains
outside the galaxy for longer and travels farther in MrAGN galaxies
than in NoAGN galaxies, and higher mass halos have the longest- and
farthest-traveling recycled gas. This is because gas launched from
lower mass galaxies is much more likely to be expelled from the halo.

Finally, \citet{Weinberger2017} implemented a prescription for AGN
feedback into AREPO, the moving mesh MHD code
\citep{Springel2010}. They included thermal feedback for
high-Eddington ratio black hole accretion and kinetic feedback for
low-Eddington ratio accretion. \citet{Pillepich2017} adopted this
dual-mode feedback prescription and also implemented an improved model
for galactic-scale, star formation-driven, kinetic winds in the
\textit{IllustrisTNG} simulation. They found that the interplay of the
new AGN feedback prescription and the new galactic wind scalings
resulted in realistic elliptical galaxies, with which we are in
qualitative agreement, but it is unclear if they drive the powerful
outflows associated with observed quasars.

Our NoAGN model is in broad agreement with several of the studies
above which only include stellar and supernova feedback. We find
similar trends in terms of inflow rate at late times and gas
recycling. However, differences in our feedback prescriptions cause us
not to agree on every detail. While our MrAGN model produces realistic
galaxies in agreement with other studies which have included AGN
feedback, our model is still unique in that we employ mechanical
feedback at all black hole accretion rates.

\subsection{Physical Interpretation}

In this work, we have examined how the cycle of gas inflow and outflow
is affected by our model for strong mechanical and radiation-driven
AGN feedback. We have focused on 24 massive galaxies with halo masses
of $M_{\rm{vir}}\sim10^{12}-10^{13.4}M_{\odot}$ at $z=0$. For each of
these galaxies, we have runs from two different models: MrAGN and
NoAGN. The MrAGN model includes stellar feedback via UV heating,
stellar winds and supernovae, AGN feedback via momentum-driven winds
and X-ray heating, photoelectric heating, and cosmic X-ray background
heating from a meta-galactic X-ray background. The NoAGN model is
identical except that it does not include any AGN feedback. The MrAGN
model has been shown to produce realistic galaxy properties for
massive galaxies \citep{Choi2015, Choi2016, Hirschmann2017}.

We set out, in part, to answer several questions about the gas cycle
in our suite of galaxies, which we will discuss here.

\subsubsection{What are the histories of inflow and outflow like for these galaxies?}

The inflow and outflow histories for both MrAGN and NoAGN galaxies are
dependent upon halo mass (Figures \ref{flowplot_163},
\ref{flowplot_329}, \ref{flowplot_501}, \ref{inflow_rate},
\ref{outflow_rate}). As demonstrated by our case studies, as well as
our ensemble plots, the inflow rate at both halo and galactic scales
is halo mass dependent and is naturally due to the depth of the
galaxy's potential well. The outflow rates also appear to be partially
mass-dependent, especially at halo scales. Outflow rarely overtakes
inflow at either scale of interest, except at late times in massive
halos (like m0163) where inflow and outflow reach rough equilibrium
due to the high central gas density. The outflows driven by AGN activity appear to correlate with AGN luminosity in a way qualitatively similar to winds in the universe (Figure \ref{outflow_rate_lbol}).

In MrAGN galaxies, halo mass governs not only the cosmological inflow
rate as for NoAGN galaxies, but also the effectiveness of AGN
feedback. The outflow rates at $\rgal$ for MrAGN galaxies are
often comparable to those of NoAGN galaxies, except for when the AGN
turns on, resulting in a spike in the outflow rate. From $z\sim$ 2 to
$z\sim 1$, the outflow rate at $\rhalo$ is elevated in MrAGN
galaxies relative to NoAGN galaxies, as material driven by AGN
feedback travels farther than material driven by stellar and supernova
feedback. We also see inflow suppressed at both radii, significantly
and for almost all galaxies at $\rgal$; after the AGN turns on,
inflow and outflow tend to track each other (Figure
\ref{loading_various}, top row). The effect is more subtle and mainly
significant for lower mass galaxies at $\rhalo$.

Both outflow enhancement and inflow suppression are seen first in
higher mass galaxies at high redshift before manifesting in lower mass
galaxies at later times, a result which corresponds to the phenomenon
of ``downsizing'', or anti-hierarchical black hole growth
\citep{Hirschmann2012, Hirschmann2014}. In our simulations this is a
result of the more massive halos receiving seed black holes at earlier
times when the gas density in the black hole's vicinity is very high,
resulting in massive black holes at higher redshift feeding
voraciously. This in turn results in more massive galaxies being
affected by AGN feedback earlier. This is in agreement with many observational studies of the AGN population which conclude that the number density
of the most luminous AGN peaks at high redshift, while less luminous
AGN have almost constant number density and are more prevalent than
high luminosity AGN at late times \citep{Cristiani2004, Croom2004,
  Matute2006}.

We also want to emphasize the increasing importance of so-called
``preventative'' feedback as we move to lower mass halos. While
``ejective'' feedback is important for removing gas and bouts of
ejective feedback are seen in halos of all masses, the inflow of gas,
both new and recycled, is far more suppressed in low mass halos than
high mass halos. By $z=1$-0, MrAGN galaxies have inflow at galactic
scales suppressed by as much as $\sim1.6$ dex, with lower mass halos
most strongly affected. In fact, the lowest mass halos even have
inflow suppressed at \textit{halo} scales by up to $\sim 0.5$ dex. The
cumulative mass of inflowing material onto the galaxy is lower by
0.4-0.6 dex for our MrAGN case studies, while the cumulative mass
accreted onto the halo is suppressed by 0.1-0.3 dex. This is a result
of outflowing material disrupting infall of new and recycled material
by imparting momentum and energy. The infall is easier to halt in a
shallower potential well. This mass-dependent preventative feedback is
an interesting consequence of our mechanical feedback
prescription. The strong outflows and their ability to clear out gas
from the galaxy are not in and of themselves a surprise as the
feedback prescription is designed to launch these winds, but their
ability to sweep up gas that is on its way into the galaxy and in some
cases turn it around is an interesting counterpoint to the usual
theoretical method of putting in two separate feedback prescriptions,
a ``wind'' mode and a ``maintenance'' mode; they are rather two sides
of the same coin. Even if our AGN feedback model might not be perfect,
since gas-free present day galaxies (such as m0501) may not be
realistic, the ability of strong outflows to act in both ejective and
preventative ways can be considered a general result and should be
kept in mind.

\subsubsection{How Much Gas is Removed Permanently and How Much Comes Back?}

The main difference between the gas cycles in MrAGN and NoAGN galaxies
is that while NoAGN galaxies remain recycling-dominated throughout
their lives, MrAGN galaxies become ejection-dominated when their black
holes begin feeding and their AGN turn on (Figure
\ref{domination_combined}). While as much as 90\% of the material in a
given outflow episode might return to a NoAGN galaxy, this fraction is
much smaller, at most 20\%, for MrAGN galaxies. Even fractions this
high are only seen in more massive galaxies whose potential wells are
deep enough to retain some of the gas pushed out by AGN
feedback. AGN-driven outflowing gas can travel larger distances than
stellar and supernova-driven gas because it is launched with much
higher velocities and is harder to slow down and turn around (Figures
\ref{ke_stellar} and \ref{vr_stellar}). MrAGN galaxies can have as
much as 90\% of the total outflowing material that is driven out past
$\rgal$ cross the virial radius by $z=0$ (Figure
\ref{fexpel}). NoAGN galaxies tend to have much smaller fractions of
expelled material, around 1-2\%. This results in NoAGN galaxies having
much higher contributions of recycled gas to their inflow and outflow
by $z=0$, with larger contributions for larger halos (Figures \ref{domination_combined} and \ref{frec}).

Since gas in NoAGN galaxies is more likely to be recycled, each gas
particle is more likely to accrete two or more times than gas
particles in MrAGN galaxies, and the timescales of these recycling
events tend to be shorter, and the distances traveled smaller, than
for MrAGN gas particles (Figures \ref{nacc_combined},
\ref{timescalehist_combined}, \ref{timescales}). Again this seems to
be due to the larger velocities associated with outflows driven by AGN
feedback, which causes the recycling history of galaxies in the two
runs to diverge when the black hole begins to feed. After that point,
MrAGN galaxies rarely recycle outflowing material. The large distances
traveled by gas particles may have interesting implications for metal
enrichment in the hot gas halo around galaxies and in the IGM (Brennan
et al. in prep).

\subsubsection{How Are the Host Galaxies Affected?}

Galaxies in our two runs tend to have different gas morphologies by
$z=0$ as a result of the different effects of feedback. In higher mass
halos, the difference is mainly in the gas density and more diffuse
hot gas halo around MrAGN galaxies as a result of more gas being
removed from both the galaxy's inner regions and its gas halo (Figures
\ref{vel_vec_163}, \ref{vel_vec_329}). In lower mass galaxies,
however, the difference is much more stark. Lower mass MrAGN galaxies
may be left only with diffuse hot gas (or may be depleted of gas
completely) while their NoAGN counterparts develop substantial discs
of cold gas by $z=0$ (Figure \ref{vel_vec_501}). While the complete
removal of gas may be too extreme, the cold gas disc, as well as the
young stellar disc that accompanies it, are in conflict with
observations in the mass range of our sample.

Beyond morphologies, AGN feedback is needed to quench our galaxies in
order for them to resemble the observed high-mass galaxy
population. Our feedback prescription accomplishes this by removing or
heating the cold gas in the galaxy, which results in a steep decrease
in star formation (Figures \ref{flowplot_163}, \ref{flowplot_329},
\ref{flowplot_501}). In the case of lower mass galaxies, as mentioned
above, even the hot gas within $\rhalo$ is severely depleted by
feedback (in m0501, this decrease is > 2 dex). This results in overall
reduced in situ SFRs and thus smaller gas (by $\sim0.5 - 2$ dex) and
stellar (by $\sim0.2$ dex) masses in MrAGN galaxies by $z=0$ (Figure
\ref{various_mhalo}).

The effects of our implementation of AGN feedback bring our simulated
galaxies into better agreement with observed galaxies in terms of
morphology, star formation rate and stellar mass, but it may result in cold gas fractions which are too low.

\section{Summary}

Our study of two sets of 24 cosmological zoom galaxies both with and
without mechanical and radiation-driven AGN feedback has led to the
following main conclusions:
\begin{itemize}
\item Our model for radiation-pressure driven AGN feedback enhances
  galaxy-scale outflows and acts in an ejective manner. Outflow rates
  at $\rgal$ are comparable between the two runs, but outflow rates at
  $\rhalo$ can be enhanced by up to 1 dex in MrAGN galaxies at $z\sim$
  1-2. This is the result of larger outflowing gas velocities in the
  MrAGN runs, which cause a higher fraction of outflowing material at
  $\rgal$ to escape the galaxy's potential well and cross
  $\rhalo$. This fraction is as high as 80\% in MrAGN galaxies,
  compared with $\sim$ 5-10\% in NoAGN galaxies.
\item Our feedback model also suppresses inflow in MrAGN galaxies
    relative to NoAGN galaxies, effecting feedback in a preventative
    way as well, especially at the low mass end of our sample. At
    $\rgal$ the inflow rate can be suppressed by as much as 1.5 dex,
    while at $\rhalo$, inflow rates can be suppressed by up to 0.5 dex
    in lower mass galaxies. This results in an overall smaller
    cumulative inflowing mass relative to the final halo masses of
    MrAGN galaxies versus their NoAGN counterparts (see Table 1 for
    these numbers for our case study halos).
\item Our NoAGN galaxies are recycling-dominated throughout their
  lives, such that most of the material removed from the galaxy
  returns ($\sim$ 68 - 78\% for our case studies). By z=0, $\sim$ 50\% of inflowing material into NoAGN galaxies is returning recycled material. Once their black holes begin to feed, MrAGN galaxies become
  ejection-dominated, with the majority of outflowing gas never returning
  to the galaxy (only about 8 - 15\% returns in our case studies). The recycled inflowing fraction of MrAGN galaxies at z=0 can be as low as a few percent.
\item Gas that \textit{is} recycled in MrAGN galaxies tends to remain
  outside of the galaxy for longer ($\sim$ 2-3 Gyrs versus 1 Gyr) and to travel farther than recycled gas in NoAGN galaxies (up to several Mpcs). Accreted gas is more likely to undergo several
  recycling events in NoAGN galaxies than in MrAGN galaxies, and is
  more likely to remain in the galaxy (either as gas or stars) until
  $z=0$. Between $\sim$ 70 - 90\% of accreted gas remains in our NoAGN case studies at $z=0$ as compared with $\sim$ 30 - 50\% in our two MrAGN case studies that have entire histories.
\item Our model for AGN feedback associated with radiatively efficient
  black hole accretion succeeds in quenching massive galaxies over
  long timescales and keeping discs from reforming.
  \end{itemize}

There are several further avenues of study which we plan to
investigate in a series of future works.

One interesting point illustrated by our individual galaxy histories
(Figures \ref{flowplot_163}, \ref{flowplot_329}, \ref{flowplot_501})
is the weak correlation between AGN activity/outflow events and galaxy
mergers. This is different from what has been seen in some other
studies of simulations with AGN feedback, such as
\citet{Tremmel2016}. While it does sometimes appear that mergers and
outflow events may be correlated, such as for m0163, it is clear that
a major merger is not a \emph{prerequisite} for quenching in our
simulations.  It is worth noting that the green dashed lines in
Figures \ref{flowplot_163}, \ref{flowplot_329} and \ref{flowplot_501}
represent 1:10 or greater halo mass ratio mergers. The stellar or
baryonic mass ratios and the time of the eventual galaxy mergers will
be somewhat different, and this could contribute to the lack of
correlation seen. It is also possible, especially in the early
universe, to have several smaller satellites, none of which represents
a 1:10 merger, interact with the central galaxy, the collective effect
of which is disruption on par with a more major interaction. AGN
activity may also be triggered by a very rich gas supply along
filaments that does not take the form of galaxy interactions. This
lack of a strong relationship between merger events and AGN activity
is also being seen in other simulations of statistically complete
galaxy populations, such as the Magneticum simulations (Steinborn et
al., in preparation). In any event, a much closer look at how this AGN
activity correlates with merger history is reserved for a future
work (Choi et al., in preparation).

We also plan to carry out a detailed comparison with observations on
several fronts. We plan to do a more in-depth study of the kinematics
of these outflows beyond the very superficial discussion here, as well
as the kinematics of the gas within the galaxy. This will include a
theoretical study of the angular momentum, as well as a study of mock
absorption lines to compare the kinematics of our simulated winds with
those observed in the universe. We also plan to use these mock
observations to compare with observed mass loading factors (again,
only briefly touched on here), in order to see how realistic our
outflows appear (especially in the case of our more extreme blowout
events in lower mass galaxies) and how our interpretation of these
events might change from different viewing angles and at different
times.

The very large distances traveled by ejected gas (and even recycled
gas in the MrAGN runs) has interesting implications for outflows
enriching the surrounding halo and the IGM, as well as enriching other
galaxies at very large distances, something which we have not studied
here. We will present a detailed study of the impact of AGN driven
winds on metal enrichment in Brennan et al. (in prep). Finally, we
also plan to explore the impact of AGN-driven outflows on nebular
emission and absorption in different regions of a galaxy, producing
spatially resolved emission and absorption line maps to improve the
interpretation of modern integral-field spectroscopic observations
(Hirschmann et al., in prep.).

In this paper we have presented a first theoretical look at the inflow
and outflow properties in a series of cosmological zoom simulations in
the presence and absence of mechanical and radiation-driven AGN
feedback. This is the first step in a series of studies in which we
will study these properties from the perspective of an observer in an
effort to compare the gas cycle in these zoom simulations with that
which can be inferred from real galaxies.

\section*{Acknowledgments}
We are grateful to the anonymous referee for constructive comments 
that improved the paper.
RB was supported in part by HST Theory grant HST-AR-13270-A. EC was
supported by NASA through HST Cycle 23 AR-14287 grant. rss thanks the
Downsbrough family for their generous support, and gratefully
acknowledges support from the Simons Foundation through a Simons
Investigator grant. MH acknowledges financial support from the
European Research Council (ERC) via an Advanced Grant under grant
agreement no. 321323-NEOGAL. RB would like to thank David Lynch and
Mark Frost for the new \textit{Twin Peaks} and Damon Lindelof and Tom
Perrotta for the final season of \textit{The Leftovers}.

%\addcontentsline{toc}{section}{\refname}\bibliography{winds}

\bibliography{winds5}

\end{document}